\documentclass[aps,pre,eqsecnum,showpacs,draft]{revtex4} 
\usepackage{amsmath,bm}

\begin{document}
 
\title{Semiclassical approach to states near potential barrier top.}
 
\author{V.A.Benderskii} 
\affiliation {Institute of Problems of Chemical Physics, RAS \\ 142432 Moscow
Region, Chernogolovka, Russia} 
\affiliation {Lab. Spectrometrie Physique, UJF \\ BP 87, St. Martin
d'Heres, Cedex, France}
 
\author{E.V.Vetoshkin} 
\affiliation {Institute of Problems of Chemical Physics, RAS \\ 142432 Moscow
Region, Chernogolovka, Russia} 
\author{E. I. Kats} \affiliation{Laue-Langevin Institute, F-38042,
Grenoble, France} 
\affiliation{L. D. Landau Institute for Theoretical Physics, RAS, Moscow, Russia}
 
\date{\today}

\begin{abstract}
Within the framework of the instanton approach we present 
analytical results for the following model problems:
(i) particle penetration through a parabolic potential barrier,
where the instanton solution practically coincides with the exact
(quantum) one; (ii) descriptions of highly excited states in two types of anharmonic
potentials: double-well $X^4$, and decay $X^3$. For the former case
the instanton method reproduces accurately not only single well and double-well
quantization but as well a crossover region (in the contrast with the standard WKB
approach which fails to describe the crossover behavior), and for the latter
case the instanton method allows to study resonance broadening and collapse
phenomena. 
We investigate also resonance tunneling, playing a relevant role
in many semiconducting devices.
We show that in a broad region of energies the instanton approach gives
exact (quantum) results.
Applications of the method and of the results may concern
the various systems in physics, chemistry and biology exhibiting
double level behavior and resonance tunneling.

\end{abstract}

\maketitle
\section{Introduction}

Semiclassical mechanics has a long history. Surprisingly however some longstanding
problems still exist in the theory. One of them, how to describe with a sufficient
accuracy states near a potential barrier top, is the subject of our paper.
It is known that the commonly used WKB method (phase integral approach) 
\cite{LL}, \cite{HE62} is reduced to matching of wave functions for
classically allowed and forbidden regions. The procedure technically 
works for linear (or first order)
turning points, 
and can be
relatively simply performed only for one dimensional problems.
However 1d problems are not of great physical importance, not only
since reduced dimensionality which does not allow to model many
relevant experimental situations but as well, at least partially, since
1d quantum mechanical problems can be rather easily solved numerically.
Unfortunately efficency and accuracy of direct quantum mechanical numerical methods 
is rapidly degraded for multidimensional systems, possessing
many degrees of freedom, because of the extraordinary amounts of computational work
required to perform these studies.
Furthermore, an extention of the WKB procedure to multidimensional
systems encounters
fundamental difficulties because of still unsolved matching problem
for the multidimensional WKB solutions which become singular on
caustic lines separating manifolds in phase space with real and imaginary
momenta for each among $N$ coordinates. Since the amount of these
domains increases like $N!$, it is a tremendous task for $N>2$. 
And after many decades of efforts, a complete and unifying descriptions of
multidimensional WKB solutions is still not available.

The problem was first addresses a long ago, and some attempts to overcome
the difficulties of the WKB approach and to improve
the accuracy of the method have been performed quiet successfully.
Note for example \cite{PK} where the authors have included into the standard
WKB method additionally a special type of trajectories on the
complex phase plane, along which the semiclassical motion is
described by the Weber functions (see also \cite{MH96}). However the choice of these
additional special trajectories (which one has to include
to improve the accuracy of the WKB method near the
barrier top) depends on the detail form of the potential far from
the top, and therefore for each particular case the non-universal
procedure should be perform from the very beginning (see also
more recent publications \cite{PJ98}, where the authors
use some distortion of Stokes diagrams, or \cite{RG02}, where time dependent
quantum mechanical calculations for anharmonic and double-well
oscillators have been performed).

Thus evidently there is some need for different from WKB semiclassical approach. 
One of the alternative to WKB semiclassical formalism
is so-called extreme tunneling trajectory or
instanton \cite{pol}, \cite{CO}, \cite{BM}, 
could be very effective for calculating globally uniform
(i.e. without any singularity) wave function of the ground state.
It allows to find semiclassical wave functions for a very
wide class of potentials with arbitrary
combinations of the first and of the second order turning points. 
The method recently was adapted as well for
the description of low - energy 
excited states \cite{BV99,BV00}. 
One of the main advantage of the instanton approach
that it can be readily extended for multidimensional systems 
using perturbative techniques (see \cite{PIA} and references
therein).

However before to investigate multidimensional problems,
one has to study 1d potentials and 1d problems which cannot
be solved accurately by the standard WKB method. It is our main
concern in the paper.
The generalization
of the instanton procedure for highly excited states is not straightforward
at all, 
and required additional
analyses. We consider only
few relatively simple examples, but  such analyses are useful
for gaining insight into more complex systems for which even approximate
theoretical results are not available. 
 
In fact for many interesting physical problems
high accuracy calculations are out
of reach by the standard WKB method but as we will see in a little while the instanton approach
is afford to overcome the difficulties of the WKB procedure.
Since this fact has largely gone unnoticed in the previous studies,
we found it worthwhile to present the investigation of a few simple
examples in a short and explicit form, and also to point out
its practical usability. And besides the study (apart from the aim
to illustrate the efficency of the instanton approach) is a prerequisite
for an explanation and successful description of many relevant physical
phenomena (e.g. low-temperature quantum kinetics of phase transitions, see
e.g. \cite{LK72})
where an active (reaction) path is confined effectively to one dimension.

All examples considered in our paper, related to the fundamental
problems of chemical dynamics and molecular spectroscopy
(see e.g. the monography \cite{BM} and references herein).
Symmetric or slightly asymmetric double-well potentials are
characteristic for molecules and Van der Waals complexes
with more than one stable configurations \cite{1}, \cite{2}, \cite{3},
\cite{5}.
The states of such systems close to the barrier top (theoretically
described by the instanton
approach in our paper) are most
relevant for radiationless evolution of highly excited states.
These states have a double nature (localized -
delocalized) and the nature manifests itself in the form of
wave functions which contain simultaneously the both components:
localized in one from the wells, and delocalized between the both
of the wells. 
The states close to the barrier top of decay potentials
govern of thermally activated over-barrier transition
amplitudes. For the low energy states the main reduction factor
is the tunneling exponent, while the contribution of the
highly excited states is limited by the Boltzmann factor.
Our instanton calculations
demonstrate that there
is no sharp boundary between quasistationary and delocalized
states.
Recently two of us (V.B. and E.K.)
\cite{BK02} investigated the eigenstates of the highly asymmetric double
well potential. We have shown that quantum irreversibility phenomena
occur when the spacing between neighbouring levels of a deeper
well becomes smaller than the typical transition matrix element.
Obviously this criterion can be also applied to the states near the
barrier top. Note that for the low energy states the asymmetry providing
irreversible behavior should be very large, whereas for the states
near the barrier top, the condition to have the ergodic behavior is not so
severe, it is sufficient that the asymmetry of the potential
is comparable to the barrier height.

Our paper has the following structure. Section II contains basic equations of the instanton
method necessary for our investigation. As an illustration of the method we
consider a touchstone quantum mechanical problem - penetration of a particle
through parabolic potential barrier. In this case the instanton solutions
which are the asymptotics of the Weber equation are exact.
Section III is devoted to the
investigation of highly excited states in a double - well potential.
For sake of concreteness and simplicity we study a quartic anharmonic $X^4$ potential.
The instanton approach allows us to reproduce accurately not only asymptotic
behavior but also a crossover region from single well to double wells quantization.
In the section IV analogous problem for $X^3$ anharmonic potential
is studied. 
Section V is devoted to
so-called resonance tunneling phenomena, interesting not only
in its own right but as well playing a relevant role
in many semiconducting double barrier structures.
Finally, in the section VI we discussed the results and conclude.
In the Appendix to the paper
we compute so-called connection matrices, which provide a very efficient
method of finding semiclassical solutions to the Schrodinger
equation in potentials having several turning points.
It is important and significant to know the connection matrices.
This is not only in itself but also for developing good analytical
approximation. Those readers who are not very interested in
mathematical derivation can skip the Appendix and find
all the results in the main body of the paper.

\section{Penetration through parabolic potential barrier.}

\subsection{Instanton approach.} 

Let us remind for the sake of conveniency the main ideas of the instanton approach.
The first step of the approach derived in \cite{pol} and
\cite{CO} is so-called Wick rotation of
a phase space corresponding to a transformation to imaginary time $ t \to i t$.
At the transformation potential and kinetic energy change their signs,
and Lagrangian is replaced by Hamiltonian in the classical equation
of motion. After this Wick rotation the standard oscillating WKB wave functions
are transformed into decaying exponentially functions which are vanished
at $ X \to \pm \infty $. Following to \cite{BV99}, \cite{BV00} we will use
slightly different formulation of the instanton method, from the very
beginning assuming exponentially decaying real-valued wave functions.
Thus taking into account that the wave functions of  bound states can be chosen as real quantities,
one can look for a solution to the Schrodinger equation in the form
\begin{eqnarray}
\label{bk1}
\Psi = \exp \left (-\gamma \sigma (X) \right ) \, ,
\end{eqnarray}
where $\gamma $ is semiclassical parameter($\gamma \equiv m \Omega _0 a_0^2/\hbar $,
where $m$ is a mass of a particle, $a_0$ is a characteristic length of the
problem, e.g. the tunneling distance, $\Omega _0$ is a characteristic
frequency, e.g. the oscillation frequency around the potential minimum, and 
further we will set $\hbar = 1$ measuring energies in the units of frequency)
which is assumed to be sufficiently large, 
and $\sigma $ can be called an action and for this function the first order
differential equation of the Ricatti type should hold
\begin{eqnarray}
\label{bk2}
\gamma ^2\left [-\frac{1}{2}\left (\frac{d\sigma }{dX}\right )^2 + V(X)\right ]
+ \gamma \left [\frac{1}{2}\frac{d^2\sigma }{d X^2} - E\right ] = 0
\, ,
\end{eqnarray}
where $V(X)$ is a potential, and $\epsilon $ gives particle eigen states (energies).
Here and below we use dimensionless variables (for the energy $\epsilon  = E/\Omega _0$,
for the potential $V = U/\gamma \Omega _0$, and for the coordinate $X= x/a_0$, where
$E$ and $U$ are corresponding dimensional values for the energy and for the potential). 
We believe that $\gamma \gg 1$ and therefore $\sigma (X)$ can be expanded
into the asymptotic series
\begin{eqnarray}
\label{bk3}
\sigma (X) = W(X) + \gamma ^{-1}W_1(X) + \gamma ^{-2} W_2(X) + ....
\, .
\end{eqnarray}
The first and the second over $\gamma ^{-1}$ order terms become identically zero,
if the time independent Hamilton-Jacobi equation (HJE) and so-called
transport equation (TE) are satisfied
\begin{eqnarray}
\label{bk4}
\frac{1}{2} \left (\frac{d W}{dX}\right )^2 = V(X)
\, ,
\end{eqnarray}
and
\begin{eqnarray}
\label{bk5}
\frac{dW}{dX}\frac{dA}{dX} + \frac{1}{2}\frac{d^2W}{dX^2}A = \epsilon A
\, ,
\end{eqnarray}
where
\begin{eqnarray}
\label{bk6}
A(X) \equiv \exp(-W_1(X))
\, .
\end{eqnarray}
The essential advantage of the instanton method in comparison
to a standard WKB is that in the former approach HJE is solved
at $E=0$, and therefore the classically allowed regions disappear.
A price we should pay for the advantage
is an appearance of second order turning points (unlike WKB
method where there are only linear turning points).

It is well known that WKB wave functions are singular at the turning points,
and therefore different approximations represent the same wave function
in various domains.
The famous Stokes phenomenom \cite{HE62} is related to the distribution
of the turning points, and Stokes and anti-Stokes lines emanate
from each turning point. By the definition Stokes lines
represent the lines where the dominance of the dominant
exponential semiclassical solution to the Schrodinger equation
becomes strongest, and anti-Stokes lines, on the other
hand represent the lines across which the dominance and subdominance
of the solutions interchange. Evidently near Stokes and anti-Stokes
lines the WKB approximation does not work and should be
refine \cite{HE62}. On the contrary, since classically
accessible regions do not exist within the instanton
formalism, the Stokes lines pass continously through second order
turning points, and globally uniform real solutions to the Schrodinger
equation can be constructed using the asymptotically smooth transformation
of the instanton wave functions into the Weber functions. Just this global
uniformity is the principal advantage of the instanton method.

A clearer idea of the instanton approach is obtained by the derivation
of well known \cite{LL} quantization rules for the harmonic
oscillator ($V(X) = X^2/2$). For a given energy $\epsilon $
any solution of the Schrodinger equation can be represented as a linear
combination of the solutions of the Weber equation \cite{GR}
\begin{eqnarray}
\label{6.1}
\frac{d^2 \Psi }{d z^2} + \left (\nu + \frac{1}{2} - \frac{z^2}{4}\right )\Psi (z)
= 0 \, ,
\end{eqnarray}
where $ z \equiv X \sqrt \gamma $, and $\nu = \epsilon - 1/2 $.
The basic solutions of (\ref{6.1}) are the parabolic cylinder functions
\cite{GR}, and only the function $D_\nu (-z)$ is vanished at $z \to \infty $
for $arg z = 0$. For the $arg z = \pi $ the asymptotic behavior of this function
at $z \to \infty $ is given by \cite{GR} 
\begin{eqnarray}
\label{6.2}
D_\nu (-z) = z^\nu \exp - \left (\frac{z^2}{4} \right ) - \frac{\sqrt {2} \pi }{\Gamma (-\nu )
}\exp(i\pi \nu ) z^{-\nu - 1}\exp \left (\frac{z^2}{4}\right )
\, ,
\end{eqnarray}
It can vanish at $z \to \infty $ only at the poles of $\Gamma (-\nu )$,
and this condition gives the exact eigenvalues of the harmonic oscillator
$$
\epsilon = n + \frac{1}{2} \, .
$$
Moreover since $D_\nu (-z)$ for integer and positive $\nu $ coincide
with known harmonic oscillator eigenfunctions \cite{LL},
the instanton approach is exact for the harmonic oscillator.

\subsection{Tunneling through the harmonic barrier.}
  
For the next less trivial illustration of the instanton approach efficiency, let us apply the method 
to the problem of quantum mechanical tunneling through a parabolic potential
\begin{eqnarray}
\label{bk10}
U(x) = U_0 - \frac{m\Omega _0^2}{2} x^2
\, ,
\end{eqnarray}
where $m$ is a mass of a tunneling particle, and $\Omega _0$ is a characteristic
frequency (curvature of the potential). Besides the potential has 
a characteristic space scale $a_0$.
Using $\Omega _0$ and $a_0$ to set corresponding scales
the parabolic potential (\ref{bk10})  
can be written in the following dimensionless form
\begin{eqnarray}
\label{bk11}
V(X) = V_0 - \frac{1}{2}X^2 \, .
\end{eqnarray}
The Schrodinger equation in these variables
\begin{eqnarray}
\label{bk12}
\frac{d^2 \Psi }{d X^2} + [\gamma ^2 X^2 - \alpha \gamma ]\Psi (X)  = 0
\, ,
\end{eqnarray}
where 
\begin{eqnarray}
\label{alpha}
\alpha = 2 \frac{U_0 - E}{\Omega _0} \, ,  
\end{eqnarray}
and introduced above semiclassical parameter 
$\gamma \gg 1 $.

The Schrodinger equation (\ref{bk12}) can be transformed into the Weber equation
\cite{GR} by $\pi /4$ rotation in the complex plane $X$
$$
X = \frac{1}{\sqrt {2}\gamma } z \exp\left (\frac{i \pi }{4}\right )\, ,
$$
and therefore the solution of (\ref{bk12}) can be represented as a linear combination of
the parabolic cylinder functions $D_\nu $ \cite{GR}
\begin{eqnarray}
\label{pc}
\Psi _\nu (z) = c_1 D_\nu (z) + c_2 D_\nu (-z) \, ,
\end{eqnarray}
where $\nu = - (1/2) - i(\alpha /2)$.

At $X \to \infty $ only the transmitted wave exists with the amplitude
(i.e. transmission coefficient) $T$
\begin{eqnarray}
\label{bk13}
\Psi \simeq T \exp \frac{i \gamma X^2}{2} \, ,
\end{eqnarray}
while for $X \to -\infty $ one has the incident ($\propto \exp(-i\gamma X^2/2)$) and
the reflected wave $\propto R\exp (i\gamma X^2/2)$. By a standard 
quantum mechanical procedure
\cite{LL} the transmission $T$ and the reflection $R$ coefficients
can be found using known asymptotics of the parabolic cylinder
functions \cite{GR} at the fixed energy (i.e. at the fixed $\alpha $) and it
leads to a well known \cite{LL} expression
\begin{eqnarray}
\label{bk14}
|T|^2 = \frac{1}{1 + \exp (\pi \alpha )} \, .
\end{eqnarray}

Note that the solutions (\ref{pc}) are the exact solutions to the Schrodinger
equation in the parabolic potential (\ref{bk11}). Now let us apply
to the same problem the instanton approach shortly described above.
The solutions of HJE (\ref{bk4}) and TE (\ref{bk5}) equations
which are milestones of the method can be found strait-forward and read
\begin{eqnarray}
\label{bk15}
W = \pm i \frac{X^2}{2} \, , \, A = A_0 X^{-1/2} \exp(\pm i \alpha \ln X/2) \, ,
\end{eqnarray}
where the integration constant $A_0$ determines energy dependent phases of
the wave functions. From the comparison of (\ref{bk15}) and (\ref{pc}) one
can see that the instanton wave functions are the asymptotics of the parabolic
cylinder functions, and, therefore, since the transmission $T$ and
reflection $R$ coefficients are determined only by the asymptotic behavior,
the values of $T$ and $R$ found in the frame work of the instanton approach
coincide with the exact quantum mechanical ones at any value of the energy
(of the parameter $\alpha $).
Let us remind that the instanton and exact quantum mechanical
solutions for the harmonic oscillator also coincide for
any energy.

To accomplish this subsection and for the sake of the skeptical
reader it is worth to mention that the WKB wave functions coincide
with the exact solutions only at $\alpha \ll -1$. In the region $|\alpha | \leq 1$,
i.e. when characteristic size of the forbidden region becomes comparable with
the particle wave length, specific interference phenomena between transmitted and reflected
waves occur, and this kind of phenomena can not be reproduced in the
standard WKB approach assuming that all turning points are independent
ones.

As an illustration in Fig. 1 we show energetical ($\alpha $) dependence 
of the phase for the wave function reflected by the parabolic potential.
The exact quantum mechanical and the instanton solutions ($\phi _0$ in Fig. 1)
are indistinguishable over a broad region of energies, while
the WKB solution ($\phi _0^{WKB}$ in Fig. 1) is deviated from both of them.

\subsection{Connection matrices.}

Our analysis can be brought into a more elegant form by introducing 
so-called connection matrices.
In the instanton approach (as in any semiclassical treatment of the
scattering or of the transition processes) we are only concerned
with the asymptotic solutions and their connections on the
complex coordinate plane.
Thus it is important and significant to know the connection matrices.
These connection matrices provide a very efficient
method of finding semiclassical solutions to the Schrodinger
equation in potentials having several turning points.
This is a relevant starting point also for developing good analytical
approximation. 

It is convenient to formulate the general procedure for
calculating of the connection matrices for an arbitrary
combinations of the first and of the second order turning points.
After that the procedure can be applied to any particular problem
under investigation.
To do it technically, one has to extend the known for linear turning
points procedure \cite{HE62}. 
All necessary details of the generalization are given in the Appendix 
to the paper, and here we present only main definitions and results.
For the equation
\begin{eqnarray}
\label{p1}
\frac{d^2 \Psi }{d z^2} + \gamma ^2 q(z) \Psi (z) = 0 \, ,
\end{eqnarray}
in the semiclassical limit $\gamma \gg 1$ the Stokes and anti-Stokes
lines are determined by the following conditions, respectively
\begin{eqnarray}
\label{pp1}
Re W(z) = 0 \, ,
\end{eqnarray}
and
\begin{eqnarray}
\label{pp2}
Im W(z) = 0 \, ,
\end{eqnarray}
where
\begin{eqnarray}
\label{p2}
W(z) = \int _{z_0}^{z} \sqrt {q(z)} dz  \, ,
\end{eqnarray}
where $z_0$ is the turning point under consideration.

For the harmonic potential we have only the linear turning points for
real (at $\alpha > 0$) 
and imaginary (at $\alpha < 0$) energies.
In the Appendix we calculated all the connection matrices we need.
Thus to fully analyze the problem for the whole range of parameters,
we should know only the distributions of turning points and Stokes 
and anti-Stokes lines on the complex plane.
At real turning points ($\alpha > 0$) $X_{1 , 2} = \pm (\alpha /\gamma )^{1/2}$,
there are 4 Stokes and 4 anti-Stokes lines, and two cuts on the complex plane
(see Fig. 2).

At $\alpha \gg 1$ the connection matrix can be easily calculated as 
the direct product of the connection matrices found in the Appendix
($\hat M^-$ from (\ref{2}) and the Hermitian conjugated matrix $\hat M^+$)
and the following diagonal shift matrix
\begin{equation} 
\label{222} 
\left( 
\begin{array}{cc} 
\exp (\pi \alpha /2) & 0   \\ 
0 & \exp (-\pi \alpha /2) 
\end{array}
\right) 
\end{equation}
It leads to the transmission coefficient $T \simeq \exp(-\pi \alpha /2)$, i.e.
coinciding with (\ref{bk14}) in the limit $\alpha \gg 1$ with the exponential
accuracy. To improve the accuracy at smaller values of $\alpha $, at the
calculation of the connection matrices, one must take into account
not only the contributions from the contours going around the turning
points, but as well the additional contribution to the action from the 
closed path (with a radius $\gg |X_{1 , 2}|$)
going around the both points $X_{1 , 2}$ (see Fig.2).
The procedure changes the Stokes constant $T_3$ (on the dashed line
separating the regions 3 and 4 in Fig. 2), which becomes
$$
|T_3| = [1 + \exp(-\pi \alpha )]^{1/2}
\, ,
$$
and finally it leads to the correct transmission coefficient
$$
T = i T_3^{-1}\exp \left (-\frac{\pi \alpha }{2}\right )
\, ,
$$
which is identical to (\ref{bk14}).

In the case $\alpha < 0$, the whole picture (see Fig. 3) of the Stokes
and of the anti-Stokes lines and turning points, is turned by the angle
$\pi /2$ with respect to depicted in Fig. 2. If one bluntly takes
the point $X=0$ as the low integration limit for the action
$W^*$ (\ref{i222}), it leads to the following transmission
coefficient
$$
T = 1 - \frac{1}{2}\exp(-\pi |\alpha | )
\, ,
$$
which can be reliable (with the accuracy $\exp(-2\pi |\alpha |)$)
only for $|\alpha \gg 1$. Again as we did for $\alpha >0$ case,
to improve the accuracy we should take into account
the contribution from the path surrounding the both imaginary turning points
(this fact was noticed long ago by Pokrovskii and Khalatnikov \cite{PK}).

At the isolated linear imaginary turning point $iX_1$ the connection matrix
is found from (\ref{2})
\begin{equation} 
\label{111} 
{\tilde M}_1^+ = \left( 
\begin{array}{cc} 
1 & i\exp (-\pi |\alpha |/2)    \\ 
0 & 1
\end{array}
\right) 
\end{equation}
Analogously the Hermitian conjugated matrix ${\tilde M}_1^-$
comes from the contribution from the closed path 
surrounding $- i X_1$. These contours provide only the
amplitude of the dominant (exponentially increasing)
wave. However the accuracy is not enough to find
the amplitude of the corresponding subdominant solution
(exponentially decaying wave function), and thus
it leads wrongly to the transmission coefficient $T=1$.
To improve the accuracy and to find correctly $T$, one
should include to the procedure the connection matrix for
the isolated second order turning point (which is
in this particular example, the maximum of the potential).
Using (\ref{i1}) we can find explicitely
this matrix
\begin{equation} 
\label{333} 
{\tilde M}_2 = \left( 
\begin{array}{cc} 
(1 + \exp (-\pi |\alpha |))^{1/2} & i \exp(-\pi |\alpha |/2)   \\ 
-i \exp(-\pi |\alpha |/2) & (1 + \exp (-\pi |\alpha |)^{1/2} 
\end{array}
\right) 
\end{equation}
In principle the same kind of calculations might be performed in
the adiabatic perturbation theory (which employs in fact
the Plank constant smallness equivalent to $\gamma \gg 1$).
Note for example the paper \cite{DY62} where 
the contributions
from the contours surrounding turning points 
(analogous to
presented above) 
have been taking into account.
It seems very plausible that on this way one will be able to combine
the instanton approach and the adiabatic perturbation theory, however,
this issue is beyond the scope of our paper and will be discussed
elsewhere.
\section{Highly excited states in double - well potential}

Literally, the instanton approach described in the previous section, is valid for the states with
characteristic energies which are small in comparison to the barrier height. However as we will show
in
this section the instanton method works pretty good for the energy states near the barrier top $V_0$. As
an illustration let us consider symmetric double well potential (quartic anharmonic $X^4$ potential)
\begin{eqnarray} 
\label{bbk1} 
V_0 - V(X) =\frac{1}{2}X^2(1-X^2) \, .
\end{eqnarray}                                                     
The Schrodinger equation with the potential (\ref{bbk1}) can be rewritten
in dimensionless variables in the following form most convenient
for the application of the instanton approach
\begin{eqnarray}
\label{bbk2}
\frac{d^2 \Psi }{d X^2} + [2 \gamma ^2 (V_0 - V(X)) - \alpha \gamma ]\Psi (X)  = 0
\, .
\end{eqnarray}
Furthermore HJE and TE instanton equations are correspondingly
\begin{eqnarray}
\label{bbk3}
\frac{1}{2} \left ( \frac{dW}{dX}\right )^2 = V_0 - V(X)
\, ,
\end{eqnarray}
and
\begin{eqnarray}
\label{bbk4}
\frac{dW}{dX}\frac{dA}{dX} + \frac{1}{2}\left (\frac{d^2W}{dX^2} + i \alpha \right )A = 0
\, .
\end{eqnarray}
Formal solutions to this set of equations (\ref{bbk3} - \ref{bbk4}) are
even and odd instanton wave functions
\begin{eqnarray}
\label{bbk5}
\Psi ^{\pm }_I = A_{\pm }(X) \exp (i \gamma W_{\pm }(X))
\, ,
\end{eqnarray}
where the action $W_{\pm }$ (solution of HJE) should be determined from
\begin{eqnarray}
\label{bbk6}
\frac{d W_{\pm }}{d X} = \pm \sqrt {2(V_0 - V(X))}
\, ,
\end{eqnarray}
and the amplitude (pre-factor) in own turn reads as
\begin{eqnarray}
\label{bbk7}
A_{\pm } = \left |\frac{d W_{\pm }}{d X}\right |^{-1/2} \exp \left (-i
\alpha \int \left (\frac{d W_{\pm }}{d X}\right )^{-1} d X \right )
\, .
\end{eqnarray}
The quantization rules \cite{LL} are related to continuous matching
of the solutions at the turning points (the second order turning point $X=0$,
and the linear turning points $X = \pm 1 $ for $\alpha > 0$ and $X = \pm i $ for
$\alpha < 0 $). The crucial advantage of the instanton solutions
(\ref{bbk5}) is due to the fact that these functions have no singularities
at all inside the barrier, since the corresponding exponents are pure imaginary
ones in the classically accessible regions (unlike WKB solutions). Besides
the general form of the instanton wave functions does not depend noticeably
on whether $E < V_0$ or $E > V_0$. This advantage allows us to include
the instanton wave functions into the basis of globally uniform functions
diagonalizing the Hamiltonian even for highly excited states.

The described above general procedure for searching instanton solutions to the Schrodinger
equation with the model potential (\ref{bbk1}) has a tricky point, what is the motivation for
presenting in some details below the explicit procedure for the search, and
new results will be emanated from our investigation.
The procedure includes several steps:
\begin{itemize}
\item
Near the second order turning point one can use the exact solution to the Schrodinger
equation (\ref{bk13}) with $c_1 = \pm c_2$ for the even and odd solutions respectively.
For $|X| \gg 1$ from (\ref{bk13}) and known asymptotics of the parabolic cylinder
functions \cite{GR}
\begin{eqnarray}
\label{bbk8}
\Psi (X) = \frac{c_1}{\sqrt X} \left [ \frac{\exp (if(X))}{\Gamma ((1 - i\alpha )/4)} +
 \frac{\exp (- if(X))}{\Gamma ((1 + i \alpha )/4)} \right ]
\, ,
\end{eqnarray}
where
\begin{eqnarray}
\label{bbk9}
c_1 = - \frac{2\pi }{\Gamma ((3 + i \alpha )/4)}\exp \left (-\frac{\pi \alpha }{8} \right )
2^{-i\alpha /4}(2\gamma )^{-1/4}
\, ,
\end{eqnarray}
and
\begin{eqnarray}
\label{bbk10}
f(X) = \frac{\gamma }{2} X^2 - \frac{\alpha }{2} ln X - \frac{\alpha }{4} ln \gamma
- \frac{\pi }{8}
\, .
\end{eqnarray}
To have correct even and odd linear combinations conforming to (\ref{bbk5})
\begin{eqnarray}
\label{bbk11}
c_{\pm } = c_1 \frac{\exp (\pm if_1)}{\Gamma ((1 \pm i \alpha /\sqrt 2)/4)}
\, ,
\end{eqnarray}
where $f_1 = (\alpha \, ln \gamma )/4 + \pi /8$.
\item
Near the linear turning point $X = \pm 1$, the Schrodinger
equation is reduced to the Airy equation \cite{GR}
\begin{eqnarray}
\label{bbk12}
\frac{d^2\Psi }{d y^2} - y \Psi (y) = 0
\, ,
\end{eqnarray}
where at $X < 0$
\begin{eqnarray}
\label{bbk13}
y = \gamma ^{2/3}\left |X + 1 + \frac{\alpha }{\gamma }\right |
\, .
\end{eqnarray}
The solution vanishing at $y \to \infty $ is \cite{GR}
\begin{eqnarray}
\label{bbk14}
\Psi (y) = |y|^{-1/4} \sin \left (\frac{2}{3}|y|^{3/2} + \frac{\pi }{4}\right )
\, .
\end{eqnarray}
Continuing this solution into the regions $(X \pm 1)\sqrt {2\gamma } \gg 1$
and sewing there with (\ref{bbk8}) we come to
\begin{eqnarray}
\label{bbk15}
\frac{c_+}{c_-} = \exp \left (- i 2 \gamma W^* + i \frac{3\pi }{2} \right )
\, ,
\end{eqnarray}
where $W^*$ is the energy dependent action in the interval $[X=0 \, , \, X=1]$.
\item
Comparing (\ref{bbk15}) and (\ref{bbk11}) we find the quantization rules which read
for the even states as
\begin{eqnarray}
\label{bbk16}
\frac{\Gamma \left ((1 + i\alpha)/4\right )}{\Gamma \left ((1 - i\alpha )/4 \right )} =
\exp \left (-2 i \gamma W^* - i \frac{3\pi }{2}\right )
\, ,
\end{eqnarray}
and for the odd states as
\begin{eqnarray}
\label{bbk17}
\frac{\Gamma \left ((3 + i\alpha)/4\right )}{\Gamma \left ((3 - i\alpha )/4 \right )} =
\exp \left (-2 i \gamma W^* - i \pi \right )
\, .
\end{eqnarray}
\item
Finally from (\ref{bbk16}), (\ref{bbk17}) we come to the quantization rule
which can be written in the single form for the both, even and odd, states 
\begin{eqnarray}
\label{bbk18}
2\gamma W^* + 2 \phi (\alpha ) \equiv
\left ( 
\begin{array}{cc} 
(5\pi /4) + 2 \pi n - \arctan (th\left (\pi \alpha /4\right )) \\ 
(3\pi /4) + 2 \pi n - 2\arctan (th\left (\pi \alpha /4\right ))  
\end{array} 
\right )  \, .
\end{eqnarray}
\end{itemize}
The relation (\ref{bbk18}) is the quantization rule we looked for, and which
allows us now to pick the fruits of the instanton method.
For the highly excited states, i.e. $\alpha \ll -1$ one can find from
(\ref{bbk18})
$$
2 \gamma W^* + 2 \phi (\alpha ) = \pi \left (n + \frac{1}{2}\right ) \, ,
$$
where $n$ is an integer number. For the low-energy states $\alpha \gg 1$
(\ref{bbk18}) reproduces the known quantization rule
$$
\gamma W^*_L = \pi \left (n + \frac{1}{2}\right ) \pm \frac{1}{2}\exp \left (-\frac{\pi \alpha }{4} \right )
\, ,
$$
where $W^*_L$ is the action in the classically admissible
region between linear turning point in the left well, namely
\begin{eqnarray}
\label{phi}
\gamma W^*_L = \gamma W^* + \phi (\alpha ) \, .
\end{eqnarray}
Note the essential advantage of the instanton quantization rule (\ref{bbk18})
with respect to the traditional WKB formalizm, where the quantization
rules in tunneling and over-barrier regions are completely different \cite{LL}.
The instanton approach gives the single quantization rule (\ref{bbk18})
which is valid in the both regions and besides fairly accurate describes the
crossover behavior near the barrier top, where periodic orbits localized at separate
wells transform into a common figure height orbit enclosing the both wells.

We illustrate the results of this section in the Fig. 4, where the universal 
dependence of the eigen  values (in the symmetric double - well potential
(\ref{bbk1}) on $\alpha $ is plotted. For the sake of comparison we presented in the
same figure the eigen  values found by the conventional WKB procedure and
by the exact quantum mechanical computation.
It is clear from the figure that the WKB method errors are maximal
in the region of small $|\alpha |$, since the oscillation period diverges logarithmically
in this region (the particle spends infinitely long time
near the second order turning point). On the contrary, the errors of the
instanton approach are minimal near the barrier top (small $|\alpha |$).

As we already mentioned at the beginning of this section, near 
the potential minimum, the instanton approach also works very
accurate. Generally speaking the instanton solutions are always
correct when the deviation from the corresponding minimum exceeds
characteristic zero - point amplitudes. Mathematically the
accuracy of the instanton approach is based on the transformation
of the semiclassical solutions into the harmonic oscillator eigenfunctions
(what ensures as well the correct normalization of the
instanton wave functions). Therefore it is naturally
to expect that the instanton method works very
accurate near the barrier top and near the potential
minimum. On the contrary, the intermediate region, where
anharmonic shape of the potential is relevant, one should
expect a poor accuracy of the instanton method.
Fortunately it turns out that the mathematical
nature of the problem is on our side, and the instanton
approach has a reasonable accuracy (of the order of the
accuracy of the WKB method) even in this region.
The fact is that the instanton wave functions are exact
not only in zero but in the first order over anharmonic
corrections to the potential approximation.
It can be shown using anharmonic perturbative procedure
proposed by Avilov and Iordanskii for WKB functions
\cite{AI} and generalized in \cite{bender} for
the instanton wave functions. 

Besides, what is more relevant for practical computations, the
instanton wave functions (unlike WKB ones) are continuous
near its ''own'' minimum. Numerical estimations show
that in the intermediate energetical region,
the accuracy of the instanton wave functions is of the
order of $5 - 10 \% $. 

To accomplish the section we present the connection matrices needed
to find semiclassical solutions to the Schrodinger
equation in the double - well potential.
Analogously to the results of the Section II, the connection
matrix for the instanton solutions is the product of the connection
matrices (\ref{2}) for the linear turning points, and the connection matrix
for the second order turning point, which is for the case under consideration,
the maximum of the double - well potential. 
Using (\ref{i1}) one can find for this latter connection matrix
\begin{eqnarray} 
\label{bb88}
\left ( 
\begin{array}{ll} 
2(\exp(\pi \alpha /2) + (1 + \exp(\pi \alpha ))^{1/2} \cos (2\gamma W^*)) \\ 
(1 + \exp(\pi \alpha ))^{1/2} \sin (2\gamma W^*) \\ 
\end{array}
\right .
\end{eqnarray}
\begin{eqnarray}
\nonumber
\left .
\begin{array}{rr} 
-(1 + \exp(\pi \alpha ))^{1/2} \sin (2\gamma W^*) \\ 
(1/2)(-\exp(\pi \alpha /2) + (1 + \exp(\pi \alpha ))^{1/2} \cos (2\gamma W^*)) 
\end{array} 
\right ) 
\end{eqnarray}

It is worth to note that the reflected wave near the barrier top
acquires non trivial phase factor. The phenomenom is related to
interference of incident, reflected and transmitted waves, and thus the
phase has some geometrical meaning, like famous Berry phase \cite{BE}.
The geometrical origin of the phase manifests itself more
clearly if we remind that the semiclassical phase factor is
determined by the probability density flow through the barrier
$$
J = i \Psi ^* \frac{d \Psi }{d X} \, .
$$
One can look to this phase factor from a slightly different point of view,
since tunneling results in the phase shift related to the change
of eigenvalues.
The quantization rules (\ref{bbk18}), (\ref{phi}) can be rewritten
in the following form, which is the definition of eigenvalues $\epsilon _n$:
$$
\epsilon _n = n + \frac{1}{2} + \chi _n \, ,
$$
where $n$ is integer numbers numerating eigenvalues, and $\chi _n$ is determined
by an exponentially small phase shift due to the existence
of the barrier between two wells. The phase
shift $\chi _n$ has the same functional form (and physical
meaning) as the geometrical phase factor (appearing due to interference
phenomena) which a quantum mechanical wave function acquires upon a cyclic evolution
\cite{BE}, \cite{WI84}, \cite{KA85}.

\section{Decay potential}

In this section we present theoretical studies of highly excited states in a decay
potential, which for definiteness will be chosen as anharmonic $X^3$ potential
\begin{eqnarray}
\label{c1}
V_0 - V(X) = \frac{1}{2}X^2(1-X)
\, .
\end{eqnarray}
As a first (but compulsory) step we investigate the low - lying tunneling
states.
\subsection{Tunneling decay of metastable states.} 
We start from this simple case to pick first the low-hanging fruits,
i.e. to describes states
\begin{eqnarray}
\label{c2}
V_0 \gg \epsilon _n \gg V(X \to \infty )
\, ,
\end{eqnarray}
what means that a local minimum is separated from continuum
spectrum by a high energetical barrier, and therefore the
quasistationary states $\epsilon _n$ are characterized by  ''good''
quantum numbers $n$.
Note that a generic decay potential, depicted in Fig. 5, is determined by the positions
of the barrier top $X_0$ and three turning points $-X_1 \, , \, X=0 \, , \,
+ X_2$, and near these points
\begin{eqnarray}
\label{c3}
V(X) =
\left \{ 
\begin{array}{cccc} 
V_0 - \left (dV/dX\right )_{X=-X_1}(X+X_1) \, , \, |X + X_1| \to 0 \\ 
\frac{1}{2}X^2 \, , \, |X| \to 0 \\
V_0 - \frac{1}{2} (X - X_0)^2 \, , \, |X - X_0| \to 0 \\
-\left (dV/dX\right )_{X=X_2}(X - X_2) \, , \, |X - X_2| \to 0   
\end{array} 
\right \}  \, .
\end{eqnarray}
The potential (\ref{c1}) is just a particular example of the generic decay
potential (\ref{c3}) ($X_1= 1/3 \, , \, X_0 = 2/3 \, , \, X_2 = 1$, and $V_0 = 2/27$)
which we will use only for the sake of
concreteness and 
explicit illustration, all presented below results are equally valid for the
generic potential. As a note of caution we should also
remark that in the instanton approach to this problem
we always have deal with the only two turning points.
For the low energy states the points are $X_2$ and the potential minimum
$X=0$, for the high energy states the points are $-X_1$ and the potential
maximum $X_0$.

According to (\ref{c1}) there are no turning points at $X > X_2$, and
at $X \gg X_2$ the potential can be considered as a constant, and
therefore the wave functions asymptotically at $X \gg X_2$ should coincide
with plane waves. Furthermore near the linear turning point $X = 1$
the Schrodinger equation with $X^3$ anharmonic potential (\ref{c1})
is reduced to the Airy equation (\ref{bbk12}), whose solutions
are linear combinations of the indices $\pm 1/3$ Bessel functions
at real (for $X<1$) and imaginary (for $X>1$) values of the arguments
\begin{eqnarray}
\label{c4}
\Psi (u) = \sqrt {u} \left [B_+ I_{1/3}\left (\frac{2 u^{3/2}}{3}\right )
+ B_- I_{-1/3}\left (\frac{2u^{3/2}}{3}\right ) \right ]
\, ,
\end{eqnarray}
and
\begin{eqnarray}
\label{c5}
\Psi (\zeta ) = \sqrt {\zeta } \left [-B_+ J_{1/3}\left (\frac{2 \zeta ^{3/2}}{3}\right )
+ B_- J_{-1/3}\left (\frac{2\zeta ^{3/2}}{3}\right ) \right ]
\, ,
\end{eqnarray}
where
\begin{eqnarray}
\label{c6}
u = (2\gamma )^{2/3}\left [1 - X - \frac{\nu + (1/2)}{\gamma } \right ]
\, ,
\end{eqnarray}
at $X<1$, and
\begin{eqnarray}
\label{c7}
\zeta  = (2\gamma )^{2/3}\left [X - 1 + \frac{\nu + (1/2)}{\gamma } \right ]
\, ,
\end{eqnarray}
at $X>1$ (remind that 
here $\nu = \epsilon _n - 1/2$).

The coefficients $B_\pm $ must be chosen such a manner that at $\zeta
\gg 1$ (\ref{c5}) gives the plane waves, and thus using
known asymptotics of Bessel functions \cite{GR}
$$
B_+ = B_- \exp\left (-i\frac{\pi }{3}\right ) \, .
$$
In the classically forbidden region $u \gg 1$ the instanton solutions of
HJE (\ref{bk4}) and TE (\ref{bk5}) matching continuously
near the turning points the quantum mechanical solutions
of the Schrodinger equation is
\begin{eqnarray}
\label{c8}
\Psi _\pm = A_\pm \exp (\pm \gamma W)
\, ,
\end{eqnarray}
where
\begin{eqnarray}
\label{c9}
A_\pm  = X^{-1/2}(1 - X)^{-1/4}\left [\frac{1 - \sqrt{1-X}}{1+\sqrt {1-X}}\right ]^{-\epsilon
_n}
\, ,
\end{eqnarray}
and
\begin{eqnarray}
\label{c10}
W = \frac{8}{15} - \frac{4}{3}(1 - X)^{3/2} + \frac{4}{5}(1 - X)^{5/2}
\, .
\end{eqnarray}
The wave functions of the quasistationary states, we are looking for,
are the linear combinations of the instanton solutions (\ref{c8})
and the suitable coefficients in the combinations
are determined from the condition of asymptotically matching to
the parabolic cylinder (\ref{pc}) and Airy functions, what leads to the following
equation for complex eigen values $\nu $
\begin{eqnarray}
\label{c11}
-\frac{\sqrt {2\pi }}{\Gamma (-\nu )} \exp \left ( \frac{16}{15}\gamma  \right )
= i \gamma ^{\nu + (1/2)}2^{6\nu +3}
\, .
\end{eqnarray}
Since the function $\Gamma (z)$ has a simple first order pole at $z=-n$,
one can easily find the main contribution to the decay rate $\Gamma _n$
of the quasistationary state $\epsilon _n$
\begin{eqnarray}
\label{c12}
\frac{\Gamma _n}{\Omega _0} = \sqrt {\frac{2}{\pi }}\frac{\gamma ^{\nu + 1/2}2^{6\nu +3}}{n!}
\exp \left (\frac{16}{15}\gamma 
\right )
\, .
\end{eqnarray}
Note that for the ground state $n=0$ (\ref{c12}) coincides to the result found
by Caldeira and Legget \cite{CL83}.
From the other hand the decay rate is related to the current flow
\cite{LL} at $X \to +\infty $ providing the constant amplitude
of the outgoing wave
\begin{eqnarray}
\label{c13}
\frac{\Gamma _n}{\Omega _0} = \left (2i\sqrt {\gamma }\int |\Psi |^2 dX \right )^{-1}\left
[- \Psi ^*\frac{d\Psi }{d X } + \Psi \frac{d \Psi ^*}{d X}\right ]
\, .
\end{eqnarray}
Introducing into (\ref{c13}) the explicit form of the wave
functions (\ref{c5}) , (\ref{c6}) , (\ref{c7}) we get
\begin{eqnarray}
\label{c14}
\frac{\Gamma _n}{\Omega _0} = \frac{9}{8}\frac{\gamma ^{1/6}}{2^{1/3}}
|B_+|^2
\, ,
\end{eqnarray}
which is equivalent to (\ref{c13}).

The decay rate according to (\ref{c14}) depends only on the normalization
of the instanton wave function and on the amplitude of the outgoing wave.
The both characteristics are determined essentially by the
behavior of the instanton wave function in the vicinity of the turning points only.
Note however that in this approximation the instanton computation
of the decay rate (\ref{c13}) or (\ref{c14}) is satisfactory one
only for the ground state, because corrections of the order of $\gamma ^{-1}$
strongly increase with the quantum number $n$. This sins of omission can be easily
relaxed if we take into account the $X^3$ anharmonic contribution
into the potential as a perturbation
\begin{eqnarray}
\label{c15}
\frac{\Gamma _n}{\Omega _0} = \sqrt {\frac{2}{\pi }}\frac{\gamma ^{\nu + 1/2}2^{6\nu +3}}{n!}
\exp \left (\frac{16}{15}\gamma \right )\left [1 -\frac{1}{576 \gamma }(164 n^3 
 + 246 n^2 + 1216 n + 567)\right ]
\, .
\end{eqnarray}
The decay rate calculated according to (\ref{c15}) provides the same level
of the accuracy as the WKB and exact quantum mechanical computations
for $\gamma \geq 5$. Out of the regime of interest, the
instanton theory loses all pretence of predictability.

\subsection{Highly excited states for anharmonic $X^3$
potential.} 

In the Subsection IV A we calculated the decay rate of low-energy
metastable states. For this case (the states $\epsilon _n$ can be characterized by
the good quantum number $n$) period of oscillations
in the well is smaller than inverse decay rate ($\epsilon _n \gg \Gamma _n/\Omega _0$),
and $\Gamma _n$ is determined by the probability current
density flowing from the well into the classically admissible
region ($X > X_0$ for a given energy $\epsilon _n$, see Fig. 5 and notations herein)
at the condition of vanishing back-flow from this
region to the barrier. Evidently the method does not work
for highly excited states, when $\Gamma _n \geq \epsilon _n \Omega _0$. In this Subsection
we go one step further with respect to the Subsection IV A
extending the instanton approach to the decay of highly excited states.

First it is worth noting that the wave functions must be vanished
at $X \to - \infty $, and besides could be always chosen as real - valued
quantities at $X \to + \infty $. From these two conditions one can find
the relations between the instanton wave functions in the regions
$X < X_1$ and $X > X_0$ (see notations in Fig. 5), and, as a result
of these relations, to calculate the phase $\delta (\alpha )$ (counted
from the barrier top) of the standing wave in the region $X > X_0$.
It reads
\begin{eqnarray}
\label{c22}
\exp (i 2 \delta )  = - i \exp(- i 2 \gamma W^*) 
\frac{1 + \sqrt {2\pi }\exp (-\pi \alpha /4)
\exp (i2\gamma W^*) [\Gamma (1-i\alpha /2)]^{-1}}
{1 + \sqrt {2\pi }\exp (-\pi \alpha /4) \exp (-i2\gamma W^*) [\Gamma (1+i\alpha /2)]^{-1}}
\, .
\end{eqnarray}
According to the standard quantum mechanics \cite{LL} the phase (\ref{c22})
determines the scattering amplitude. Thus from (\ref{c22}) we can find
the scattering amplitude in the deep classically forbidden region, and as
a result of it to compute the eigen values in this region. To perform
the calculation we need to know the terms of the order of $\exp (-\pi |\alpha |)$
in the $\Gamma $-functions expansion (which are beyond the standard Stirling
formula) \cite{GR}
\begin{eqnarray}
\label{c23}
\Gamma \left (\frac{1 \pm i\alpha }{2} \right ) \simeq
\sqrt {2\pi }\exp \left (-\frac{\pi \alpha }{4} \pm i \phi \right ) \left
(1 - \frac{1}{2}\exp (-\pi |\alpha |)\right )
\, ,
\end{eqnarray}
where
\begin{eqnarray}
\label{c24}
\phi (\alpha ) \equiv\frac{\alpha }{2} \left [ \ln \frac{|\alpha |}{2} - 1\right ]
\, .
\end{eqnarray}
Finally from (\ref{c22}), and taking into account (\ref{c23}), (\ref{c24}) 
we find (with required exponential accuracy)
the poles of the scattering amplitude 
\begin{eqnarray}
\label{c25}
2 \gamma W_L^* = 2 \gamma W^* + \phi (\alpha ) =
\pi (2 n +1) - i \left [\frac{\pi }{4}(|\alpha | - \alpha ) + \frac{1}{2} \exp (-\pi |\alpha |)
\right ]
\, ,
\end{eqnarray}
and therefore explicitely solving (\ref{c25}),
the complex eigen values, and in particularly the decay rate for
highly excited states in the anharmonic decay potential.

The same manner as for low energy tunneling states,
for highly excited states (i.e. at $|\alpha | \gg 1$) the real part of
the eigen values $\epsilon _n$ is determined by the action along closed
trajectories in the well, whereas the imaginary part (i.e.
the decay rate $\Gamma _n$) is related to the probability
current density flow from the well to the barrier.

Using the instanton approach procedure shortly described in the Sections II
and III (for the details see \cite{BV99}, \cite{BV00}), in own turn one
can find not only eigen values but as well eigen states.
The real-valued instanton wave functions are determined
by the action $W(X_1, X)$ which is counted from the linear
turning point $X_1$
\begin{eqnarray}
\label{c26}
\Psi (X) = A(\alpha) |X - X_1|^{-1/4} \sin (\gamma W(X_1, X) + \pi /4)
\, ,
\end{eqnarray}
where the amplitude $A(\alpha )$ of the wave function acquires
maximum values at the poles of (\ref{c22}) with widths
proportional to $\Gamma _n$.
We plotted the functions $|A(\alpha )|^2$ 
in Fig. 6.

\section{Resonance tunneling.} 

The phenomenom of the electron resonance tunneling
is a familiar one \cite{BO51} and observed (see e.g. \cite{CE74}, and for more
recent references also \cite{EG91}) in semiconducting
heterostructures possessing so-called double barrier potentials (see Fig. 8).
The phenomenom manifests itself as peaks in the tunneling current at voltages
near the quasistationary states of the potential well. The physical mechanism
of the resonance tunneling can be understood as a constructive interference
between the wave reflected from the left barrier and outgoing to the left
from the well.

In the instanton method the total transmission coefficient $T$
is determined by the second order turning points of the double barrier
potential (i.e. the maxima of the potential) and according
to the procedure shortly described in the Section II, $T$ reads as
\begin{eqnarray}
\label{c16}
|T|^2 = \frac{\pi ^2 \Gamma _L \Gamma _R}{(1 - \sqrt {(1+\pi \Gamma _L)(1 + \pi \Gamma _R)})^2
+ 4\sqrt {(1 + \pi \Gamma _L)(1+ \pi \Gamma _R)}\cos ^2(\gamma W^*_R)}
\, ,
\end{eqnarray}
where we designate
\begin{eqnarray}
\label{c17}
\Gamma _{L , R}  = \frac{1}{\pi }\exp (-\pi \alpha _{L , R})
\, ,
\end{eqnarray}
as above (\ref{alpha})
\begin{eqnarray}
\label{c18}
\alpha _{L , R} = 2 \frac{U_{0 (L , R)} - E}{\Omega _{0 (L , R)}}
\, ,
\end{eqnarray}
and the action in the classically admissible region analogously to (\ref{phi}) is
\begin{eqnarray}
\label{c19}
\gamma  W_R^* = \gamma W^* - \phi (\alpha _L) - \phi (\alpha _R)
\, .
\end{eqnarray}
In the resonance region, where according to the stationary
quantization rule
$$
\gamma W_R^* = \pi \left (n + \frac{1}{2}\right )
\, ,
$$
the transmission coefficient from (\ref{c16})
is
\begin{eqnarray}
\label{c20}
|T|^2 = \frac{4 \Gamma _L \Gamma _R}{(\Gamma _L + \Gamma _R)^2}
\, ,
\end{eqnarray}
and far from the resonance
\begin{eqnarray}
\label{c21}
|T|^2 = \frac{\pi ^2 \Gamma _L \Gamma _R}{4\cos ^2(\gamma W^*_R)}
\, .
\end{eqnarray}
Thus we found the resonance amplification of the transmission.
For the symmetric case in the resonance $T=1$, and the interference
suppresses completely reflection. In the opposite case of strongly asymmetric
barriers $T$ from (\ref{c16})
is almost coincided with the transmission coefficient for the highest
barrier, and an influence of the lower barrier is suppressed by the interference.
In Fig.9 we show the energy dependence of $T$ for the symmetrical structure of the
barriers. The resonances become broader when the energy approaches
to the potential barriers top, and disappear at higher energies (above the top).
It is worthwhile to stress that the instanton solution of the resonance tunneling
problem allows us to study the phenomenom in a very broad
energetical region, including the states near the barriers tops.

Finally let us present the connection matrices for 
the found above instanton solutions. The corresponding matrix can be found as
the product of two connection matrices connecting instanton solutions
near the second order turning points (see (\ref{i1}) and (\ref{bb88})), and the
diagonal shift matrix (\ref{ii1}):
\begin{eqnarray}
\label{cc88}
\left ( 
\begin{array}{ll} 
2\pi \exp(\pi (\alpha _L + \alpha _R /4)\exp (i \gamma W^*)\{\Gamma [(1+i \alpha _L)/2]
\Gamma [(1 + i\alpha _R)/2]\}^{-1} + \exp (\pi (\alpha _L+ \alpha _R)/2 )(\exp (-i \gamma W^*) \\
- i{\sqrt {2\pi }}\exp [(3\pi (\alpha _R + \alpha _L)/8]\left (
\exp (i \gamma W^*)\Gamma ^{-1}[(1 + i \alpha _R)/2] +
\exp (-i \gamma W^*)\Gamma ^{-1} [(1 - i \alpha _L)/2] \right )
\end{array}
\right. 
\end{eqnarray}
\begin{eqnarray}
\nonumber
\left. 
\begin{array}{rr} 
i{\sqrt {2\pi }}\exp [(3\pi (\alpha _R + \alpha _L)/8]\left (
\exp (i \gamma W^*)\Gamma ^{-1} [(1 + i \alpha _L)/2] +
\exp (-i \gamma W^*)\Gamma ^{-1} [(1 - i \alpha _R)/2] \right ) \\
2\pi \exp(\pi (\alpha _L + \alpha _R /4)\exp (-i \gamma W^*)\{\Gamma [(1+i \alpha _L)/2]
\Gamma [(1 + i\alpha _R)/2]\}^{-1} + \exp (\pi (\alpha _L + \alpha _R)/2 )(\exp (i \gamma W^*)
\end{array} 
\right )  
\end{eqnarray}
Here as always we designated by $W^*$ the action between the turning
points (in this case between the second order turning points).

\section{Conclusion}
Our paper could be considered as a formal one,
in the sense that we asked theoretical questions that
most of solid state or chemical physics experimentalists
would not think to ask. However, the answering of these very basic
questions can be illuminating.

Let us sum up the results of our paper.
Within the framework of the instanton approach 
we derived accurate analytical solutions 
for a number of 1d semiclassical problems, and checked numerically
the results. As an illustration of the method we
consider a simple quantum mechanical problem - penetration of a particle
through the parabolic potential barrier. In this case the instanton solutions
which are the asymptotics of the Weber equation are exact.
The second investigated problem is related to the
description of highly excited states in a double - well potential.
For sake of concreteness and simplicity we study a quartic anharmonic $X^4$ potential.
The instanton approach enables us to reproduce accurately not only asymptotic
behavior but also a crossover region from single well to double wells 
quantization (in a contrast with the standard WKB
approach which fails to describe the crossover behavior).
The analogous problem for $X^3$ anharmonic potential
is also studied, and the instanton method allows to study resonance broadening and collapse
phenomena.
Besides we investigated  
so-called resonance tunneling phenomena, interesting not only
in its own right but as well playing a relevant role
in many semiconducting double barrier structures.
We computed as well the connection matrices, which provide a very efficient
method of finding semiclassical solutions to the Schrodinger
equation in potentials having several turning points.
It is also useful for developing good analytical
approximation. 

All examples selected in our paper to illustrate the efficency
of the instanton approach, belong to the fundamental
problems of chemical dynamics and molecular spectroscopy
(see e.g. the monography \cite{BM} and references herein).
Symmetric or slightly asymmetric double-well potentials are
characteristic for molecules and Van der Waals complexes
with more than one stable configurations \cite{1}, \cite{2}, \cite{3},
\cite{5}.
The states of such systems closed to the barrier top (theoretically
described by the instanton
approach in our paper) are not easy to investigate experimentally,
since optical transitions between these states and localized ones,
are typically inactive. However just these states are most
relevant for radiationless evolution of highly excited states.
In a certain sense these states have a double nature (localized -
delocalized) and the nature manifests itself in the form of
wave functions which contain simultaneously the both components:
localized in one from the wells, and delocalized between the both
of the wells. As a consequence of it, any initially prepared
localized state will evolve via formation and decay of these
states. Our calculations are intended to pave a way to investigate this
class of problems using the wave functions computed
within the instanton approach.

The states close to the barrier top of decay potentials
govern of thermally activated over-barrier transition
amplitudes. For the low energy states the main reduction factor
is the tunneling exponent, while the contribution of the
highly excited states is limited by the Boltzmann factor.
The energetical width of the region dominating in the total
transition rate is postulated traditionally in the transition rate theory
\cite{4} as being of the order of the temperature $T$.
However our results presented in the Section III predict
another estimation. Instanton calculations
demonstrate that the intermediate region between the quasistationary
$\Gamma \ll \Omega $ and delocalized states could be much larger
than $T$, namely of the order of $\Omega $. It means that there
is no sharp boundary between quasistationary and delocalized
states and all of the states within the interval $V^* - \Omega \, , \, V^* + T$
equally contribute to the total rate constant for the penetration
through the barrier.

One more point should be emphasised. Recently in
\cite{BK02} has been shown that quantum irreversibility phenomena
occur when the spacing between neighbouring levels of a deeper
well becomes smaller than the typical transition matrix element.
Obviously this criterion can be also applied to the states near the
barrier top. Note that for the low energy states the asymmetry providing
irreversible behavior should be very large, whereas for the states
near the barrier top, the condition to have the ergodic behavior is not so
severe, it is sufficient that the asymmetry of the potential
is comparable to the barrier height.

Applications of the method and of the results may concern
also the various systems in physics, chemistry and biology exhibiting
double level behavior and resonance tunneling.
Literally speaking in this paper we dealt with the 
microscopic Hamiltonians. However, thanks
to the rapid development of electronics and cryogenic technologies,
it has become possible to apply the same Hamiltonians to study cases where
the corresponding variables are macroscopic (e.g. magnetic flux through
a SQUID ring, or charge or spin density waves phase in
a certain one dimensional solids).
For example in the paper we studied tunneling processes
in which a system penetrates into a classically forbidden
region (a potential barrier). It is an intrinsically quantum
effect with no classical counterpart, but nevertheless
it can take place for macroscopic systems, and the tunneling
of a macroscopic variable of a macroscopic system 
(e.g. spin or charge tunneling in atomic condensates
trapped in a double-well potential \cite{PZ01}) can be
also investigated by our method.

With this background in mind our results are also intended to
clarify different subtle aspects of tunneling phenomena. An example was given
at the end of the Section III , where we found the geometrical
phase acquired by a particle tunneling through a potential barrier.
This phase can be tuned by the particle energy and by the barrier
shape, and specific interference phenomena might occur.
The observation of oscillations related to this geometrical phase
in real systems has proved challenging.
Of course since the forms of the model potentials we used are
rather specialized (and besides only 1d), we cannot discuss
the behavior for general cases with full confidence.
Nevertheless, we believe that the instanton approach
employed in the present work should be useful
in deriving valuable results for the general and multidimensional
potentials as well.

Essentially that in the instanton method, we discussed in the
paper, just observing a few classical trajectories suffices
to develop a qualitative insight for quantum behavior.
Though (as we illustrated in a number of particular examples
considered in the paper) relied upon in this way, semiclassical
instanton approach is much more than a qualitative picture. As an approximation,
the instanton method can be surprisingly precise.
Note also that results presented here
not only interested in their own right (at least in our opinion)
but they might be directly tested experimentally since there
are many systems where the model investigated in the paper 
is a reasonable model for the reality. 

The theory presented in our paper could be extended in several
directions. One very interesting question is how our quantum
mechanic instanton formulas (e.g. (\ref{c12}) - (\ref{c14})
for the tunneling rate
in the anharmonic $X^3$ decay potential) are modified by interactions
with surrounding media (see e.g. \cite{MY00}, where WKB approach
was used to study the time evolution of quantum tunneling in a thermally
fluctuating medium).
However the theoretical modelling of this case is hampered
by lack of detailed knowledge of the medium and of the interaction
with it. A more specific study might become
appropriate should suitable experimental results become available.
A simple criterion for the strength of interaction with an environment (or by other
words for the effective temperature) where roughly the crossover from
thermally activated classical to quantum mechanical decay
can be found easily by equating the corresponding
Arrhenius factor and characteristic frequency ''oscillations''
inside the barrier $\Omega _*$ (see (\ref{i2})). 

All of the potentials investigated in our paper can be considered
in a number of realistic cases as effectively resulting from
avoiding of adiabatic level crossing 
in the situation, when the adiabatic splitting
is so large that any influence of the upper adiabatic states
on the lower states can be neglected. Certainly in a general
case of an arbitrary coupling strength, this interaction
of higher and lower adiabatic states must be taken into account,
and the tunneling matrix elements are accompanied by corresponding
Franck - Condon factors arising due to violation of Born
Oppenheimer approximation.
We defer these problems
to the further, although no any doubt the instanton approach
is useful for this kind of problems as well.

\acknowledgements
The research described in
this publication was made possible in part by RFBR Grants 00-03-32938 and 00-02-11785.

\appendix

\section{}

For the standard basic WKB solutions following \cite{HE62} we will
introduce shorthand notations
\begin{eqnarray}
\label{p3}
(\circ , z) \equiv (q(z))^{-1/4} \exp (i \gamma W(z))
\, ,
\end{eqnarray}
and
\begin{eqnarray}
\label{p4}
(z , \circ ) \equiv (q(z))^{-1/4} \exp (- i \gamma W(z))
\, ,
\end{eqnarray}
The position of the turning point is designated by $\circ $
and not essential if we are looking for the solutions in the region
$|z| \gg 1$.
According to the definitions (\ref{pp1}), (\ref{pp2}) on the Stokes
lines one should add the dominant solution times a certain constant
(Stokes constant) to the subdominant (decaying) solution, while
on the anti-Stokes lines the dominant and the subdominant
solutions are exchanged. To find the Stokes constant we must match
the both solutions going around the turning point and taking into
account the cut on the complex $z$ plane (see Fig. 10).

Let us consider first the linear turning point
\begin{eqnarray}
\label{p5}
q(z) = - z \, ,
\end{eqnarray}
when the classically admissible region corresponds to $X > 0$.
For this case we have 3 Stokes lines, 3 anti-Stokes lines, one cut,
and therefore 7 different regions on the complex $z$-plane where the functions
(\ref{p3}), (\ref{p4}) should be matched, and after that 3 Stokes
constant should be determined. After not very sophisticated
but rather tedious algebra we are ending with all three Stokes constants
$$
T_1 = T_2 = T_3 = i \, ,
$$
and the connection matrix related the coefficients of the 
combinations of the basic solutions (\ref{p3}), (\ref{p4}) in the
classically are ending ($A_1 , A_2$) and in classically admissible ($A_2 ,
B_2$) regions are
\begin{equation} 
\label{1}
\left( 
\begin{array}{c} 
A_2  \\ 
B_2 
\end{array}
\right) =
{\hat M}^{-} 
\left( 
\begin{array}{c} 
A_1  \\ 
B_1 
\end{array}
\right) 
\end{equation}
where
\begin{equation} 
\label{2} 
{\hat M}^{-} =
\exp \left (- i\frac{\pi }{4}\right ) \left( 
\begin{array}{cc} 
\exp (i\pi /4 & (1/2)\exp (-i\pi /4)   \\ 
\exp (-i \pi /4) & (1/2) \exp (i\pi /4) 
\end{array}
\right) 
\end{equation}
Analogously for the other linear turning point
$q(z) = +z$, the connection matrix $\hat M^+$ is Hermitian conjugated to
$\hat M^- $. The variation of the coefficients in the region
between two independent linear turning points $z_1$ and $z_2$ are
determined by the diagonal matrix $\hat L$
\begin{equation} 
\label{ii1}
{\hat L} =  \left( 
\begin{array}{cc} 
\exp (- i\gamma W^*)  & 0   \\ 
0 & \exp (+i \gamma W^*) 
\end{array}
\right) 
\end{equation}
where
$$
W^* = \int _{z_1}^{z_2} \sqrt {q(z)} dz \, .
$$                                                                                                         
And finally for the solutions in the classically
forbidden regions $X < X_1$ and $X > X_2$ the connection matrix is
the direct matrix product of the found above matrices
$$
\hat {M} = \hat {M}^+\hat {L}\hat {M}^- \, .
$$
To generalize the procedure for the second
order turning points we should find the connection
matrices related the basic solutions to the Weber equation, namely
\begin{eqnarray}
\label{pp4}
(\circ , z ) \equiv (z)^{\nu} \exp (- z^2/4)
\, ,
\end{eqnarray}
and
\begin{eqnarray}
\label{pp5}
(z , \circ ) \equiv (z)^{-\nu - 1} \exp (z^2/4)
\, .
\end{eqnarray}
In this case we have 4 Stokes lines, 4 anti-Stokes lines, one cut, and therefore
9 different regions, where the solutions should be matched (see Fig. 11a
as an illustration). Four Stokes constants are
$$
\tilde T_1 = \tilde T_2^{-1}(\exp (i2\pi \nu ) - 1)
\, , \, \tilde T_3 = \tilde T_1^* \, , \,
\tilde T_4 = - \tilde T_2 \exp(-i2\pi \nu ) \, .
$$
And from the known asymptotic of the parabolic cylinder functions
one can get the last remaining Stokes constant
$$
\tilde T_2 = \frac{\sqrt {2 \pi }}{\Gamma (-\nu )} \, .
$$
Thus the connection matrix for an isolated second order turning
point can be represented as
\begin{equation} 
\label{i1}
\left( 
\begin{array}{cc} 
-\tilde T_2  & \cos (\pi \nu )   \\ 
\cos (\pi \nu ) & -\sin ^2(\pi \nu )/\tilde T_2 
\end{array}
\right) 
\end{equation}
For example this depending on the energy $\epsilon $ matrix
determines instanton semiclassical solutions for the harmonic
oscillator $\epsilon = \nu + 1/2$. It can be checked by explicit
calculations that for the harmonic
oscillator the same form (\ref{i1}) of the connection matrix
holds as well for WKB approach. The difference could appear
only from anharmonic terms in the potential.
However for low energy states $\epsilon /\gamma \ll 1$
anharmonic corrections are small and up to the second order
over these corrections terms the both connection matrices (instanton
and WKB) coincide.

The connection matrix describing the variation of the coefficients
at the basic solutions
(\ref{pp4}), (\ref{pp5}) between two second order turning points
$X_2^0$ and $X_3^0$ for the symmetrical double-well potential is
(compare with the analogous matrix (\ref{ii1}) for two linear turning
points)
\begin{equation} 
\label{i2}
\left( 
\begin{array}{cc} 
(n!/\sqrt {2\pi })(\Omega _0\gamma /\Omega _*)^{-\nu + 1/2} \exp (\gamma W_E^*)  & 0   \\ 
0 & (\sqrt {2\pi }/n!)(\Omega _0\gamma /\Omega _*)^{\nu + 1/2} \exp (-\gamma W_E^*)   
\end{array}
\right) 
\end{equation}
where the instanton action
\begin{eqnarray}
\label{i222}
W_E^* = \int _{X_2^0}^{X_3^0} \sqrt {2(V(X) - (\epsilon /\gamma )} dX \, ,
\end{eqnarray}
and $\Omega _*$ is the characteristic ''oscillation'' frequency in the barrier
(i.e. in the classically forbidden region).
For the asymmetric double-well potential in the region between the second
order and the linear turning points the analogous to (\ref{i2}) matrix is
\begin{equation} 
\label{i3}
\left( 
\begin{array}{cc} 
(n!/\sqrt {2\pi })^{1/2}(\Omega _0\gamma /\Omega _*)^{-(1/2)(\nu + 1/2)} \exp (\gamma W_E^*)  & 0   \\ 
0 & (\sqrt {2\pi }/n!)^{1/2}(\Omega _0\gamma /\Omega _*)^{(1/2)(\nu + 1/2)} \exp (-\gamma W_E^*)   
\end{array}
\right) 
\end{equation}
All given above matrices allow us to find any other
connection matrices we need for all particular examples considered
in the main text of the paper. Any of them can be constructed as a corresponding
product of
(\ref{1}), (\ref{2}), (\ref{ii1}), (\ref{i1}), (\ref{i2}), (\ref{i3}).
It is worth noting at the very end one general property of the connection
matrices, namely that for all bound states
the connection matrix is real-valued, and for continuum spectrum states, off-diagonal
elements of the connection matrix are complex.

Analogously for the problem of tunneling through the potential barrier
$V(X) = - (1/2)X^2$, all Stokes and anti-Stokes lines are turned by 
the angle $\pi /4$
(see Fig. 11b) with respect to the corresponding lines for
the parabolic well ($V(X) = (1/2)X^2$ considered above, see Fig. 11a).
The connection matrix for the tunneling through the barrier
\begin{equation} 
\label{33}
\left( 
\begin{array}{cc} 
(S_1  & -i \exp(\pi \alpha /2)   \\ 
i \exp(\pi \alpha /2)  & S_1^{-1} \left (\exp (\pi \alpha ) + 1 \right )   
\end{array}
\right) 
\end{equation}
where $\alpha = i(2 \nu + 1)$ and $S_1$ is the Stokes constant on
the first quadrant bisectrix (see Fig. 11b).
To find the Stokes constant $S_1$ one has to match the sum
of the incident and of the reflected waves to the solutions
of the Weber equation at $ X \to -\infty $ and to the transmitted wave
at $X \to \infty $. It leads
$$
S_1 = \frac{\sqrt {2 \pi }}{\Gamma [(1 + i\alpha )/2]}\exp (\pi \alpha /4)
\, .
$$

\newpage

\centerline{Figure Captions.}

Fig. 1

The phase of the reflected from the parabolic barrier wave:

(1) - exact quantum and instanton solutions $\phi _0$;

(2) - WKB solution $\phi _0^{WKB}$;

dashed line is the difference $\phi _0 - \phi _0^{WKB}$.

Fig. 2

The Stokes (solid) and anti-Stokes (dashed) lines for the two
real-valued turning points $X_{1 , 2}$ with the surrounding
contours $I$ and $I^\prime $. On the contour $II$ the Stokes
lines for the Airy equation asymptotically matched to the
lines for the Weber equation. The cut is depicted
by the wavy line.

Fig. 3

The same as Fig. 2 but for the case of the two pure
imaginary turning points $i X_{1 , 2}$.

Fig. 4

The dimensionless tunneling splitting $\Delta /\Omega _0$
for the anharmonic $X^4$ potential near the barrier top:

(1) - exact quantum and instanton calculations;

(2) - WKB result.

Fig. 5

The $X^3$ anharmonic decay potential.

Fig. 6

The amplitude of the wave function localized in the potential shown
in Fig. 5 (the dashed line is a non-resonant part of the amplitude,
and $\gamma = 101$).

Fig. 7

The model two barrier potential structure for the resonance tunneling.

Fig. 8

The transmission coefficient for the potential shown in Fig. 7
($\gamma = 54 $).

Fig. 9

The Stokes (solid) and anti-Stokes (dashed) lines in the vicinity of
the linear turning point $V(X) = - X$. The cut is depicted by the wavy line,
and the Stokes constant are $T_1$, $T_2$, and $T_3$.

Fig. 10

The Stokes and anti-Stokes lines in the vicinity of the second order
turning points (the same notations as in the Fig. 9):

(a) $V(X) = (1/2)X^2$ ;

(b) $V(X) = - (1/2)X^2$.


\begin{references} 
\bibitem{LL} L.D.Landau, E.M.Lifshits, Quantum Mechanics (non-relativistic
theory), Pergamon Press, New York (1965). 
\bibitem{HE62} J.Heading, An Introduction to Phase-Integral
Methods, Wiley - Interscience, London (1962).
\bibitem{PK} V.L.Pokrovskii, I.M.Khalatnikov, JETP, {\bf 13}, 1207 (1961).
\bibitem{MH96} N.T.Maintra, E.J.Heller, Phys. Rev. A, {\bf 54}, 4763 (1996).
\bibitem{PJ98} C.S.Park, M.C.Jeong, Phys. Rev. A, {\bf 58}, 3443 (1998).
\bibitem{RG02} A.K.Roy, N.Gupta, D.M.Deb, Phys. Rev. A, {\bf 65},
012109 (2002).
\bibitem{pol} A.M.Polyakov, Nucl.Phys. B, {\bf 129},
429 (1977).
\bibitem{CO} S.Coleman, Aspects of Symmetry, Cambridge University Press,
Cambridge (1985).
\bibitem{BM} V.A.Benderskii, D.E.Makarov, C.A.Wight, Chemical
Dynamics at Low Temperatures, Willey-Interscience, New York (1994).
\bibitem{BV99} V.A.Benderskii, E.V.Vetoshkin, H.P.Trommsdorf, Chem. Phys., {\bf 244},
273 (1999).
\bibitem{BV00} V.A.Benderskii, E.V.Vetoshkin, 
Chem. Phys., {\bf 257}, 203 (2000). 
\bibitem{PIA} V.A.Benderskii, E.V.Vetoshkin, L.S.Irgebaeva,
H.P.Trommsdorff, Chem. Phys., {\bf 262}, 369 (2000). 
\bibitem{LK72} I.M.Lifhsits, Yu.Kagan, JETP, {\bf 35}, 206 (1972).
\bibitem{1} M.Grifoni, P.Hanggi, Phys. Repts., {\bf 304}, 229 (1998).
\bibitem{2} J.E.Avron, E.Gordon, Phys. Rev. A, {\bf 62}, 062504 (2000).
\bibitem{3} J.Ankerhold, H.Grabert, Europhys. Lett., {\bf 47}, 285 (1999).
\bibitem{5} Y.Kaganuma, Y.Mizumoto, Phys. Rev. A, {\bf 62}, 061401(R) (2000).
\bibitem{BK02} V.A.Benderskii, E.I.Kats, Phys. Rev. E, {\bf 65}, 036217 (2002).
\bibitem{AI} V.V.Avilov, S.V.Iordanskii, JETP, {\bf 69}, 1338 (1975).
\bibitem{bender} V.A.Benderskii, E.V.Vetoshkin, L. von Laue, H.P.Trommsdorff,
Chem. Phys., {\bf 219}, 143 (1997).
\bibitem{DY62} A.M.Dykhne, JETP, {\bf 14}, 941 (1961).
\bibitem{GR} A.Erdelyi, W.Magnus, F.Oberhettinger, F.G. Tricomi,
Higher Transcendental Functions, vol.1 - vol.3, McGraw Hill,
New York (1953).
\bibitem{BE} M.Berry, Proc. Roy. Soc. London, ser. A,
{\bf 392}, 45 (1984).
\bibitem{WI84} M.Wilkinson, J.Phys. A, {\bf 17}, 3459 (1984).
\bibitem{KA85} H.Karatsuji, Progr. Theor. Phys., {\bf 74}, 439 (1985).
\bibitem{CL83} A.O.Caldeira, A.J.Legget, Ann. Phys., {\bf 149}, 374 (1983).
\bibitem{BO51} D.Bohm, Quantum Theory, Englewood Clifs, New Jersy (1951).
\bibitem{CE74} L.L.Chang, L.Esaki, R.Tsu, Appl. Phys. Lett., {\bf 24}, 593 (1974).
\bibitem{EG91} J.P.Eisenstein, L.N.Pfeiffer, K.W.West, Phys. Rev. Lett.,
{\bf 74}, 1419 (1995).
\bibitem{MY00} Sh.Matsumoto, M.Yoshimura, Phys. Rev. A, {\bf 63}, 012104 (2000).
\bibitem{4} M.Baer, Phys. Repts., {\bf 358}, 75 (2002).
\bibitem{PZ01} H.Pu, W.Zhang, P.Meystre, Phys. Rev. Lett., {\bf 87},
140405 (2001).

\end{references}
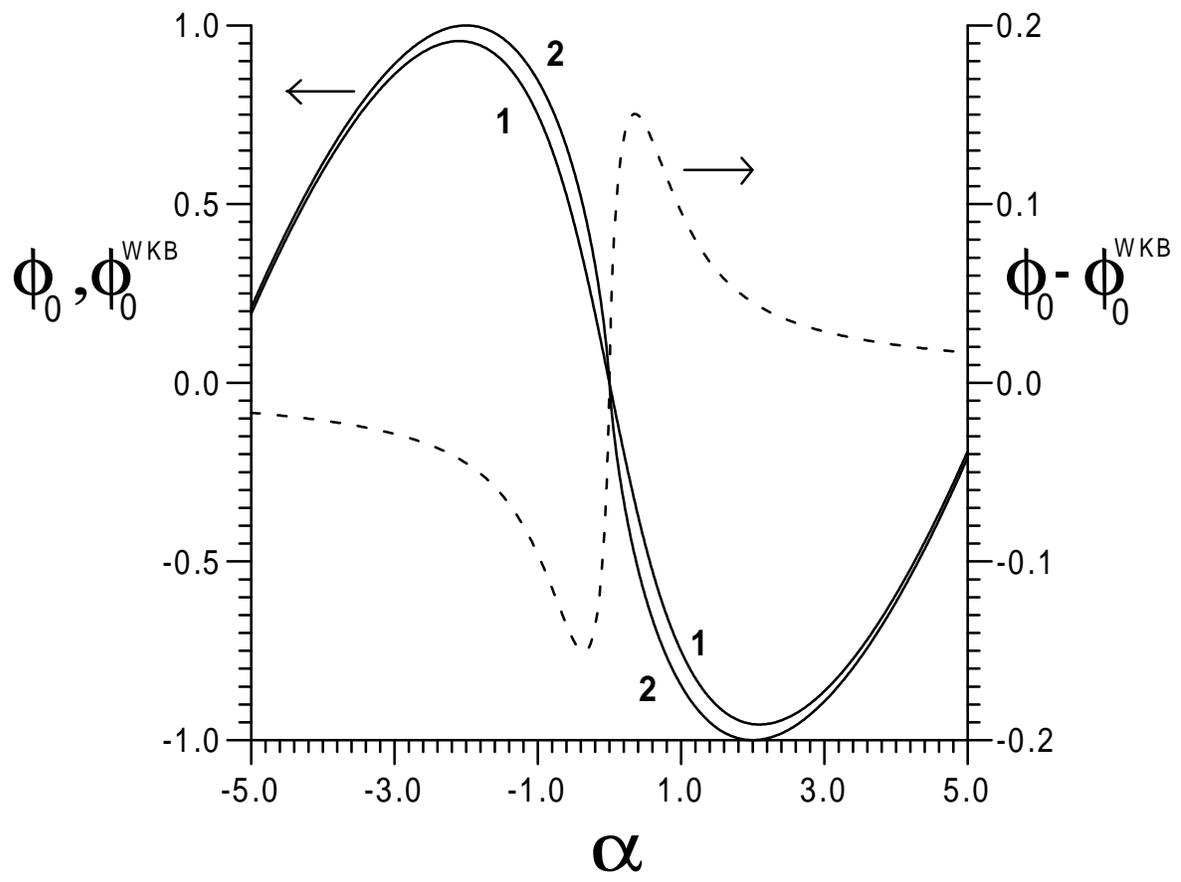
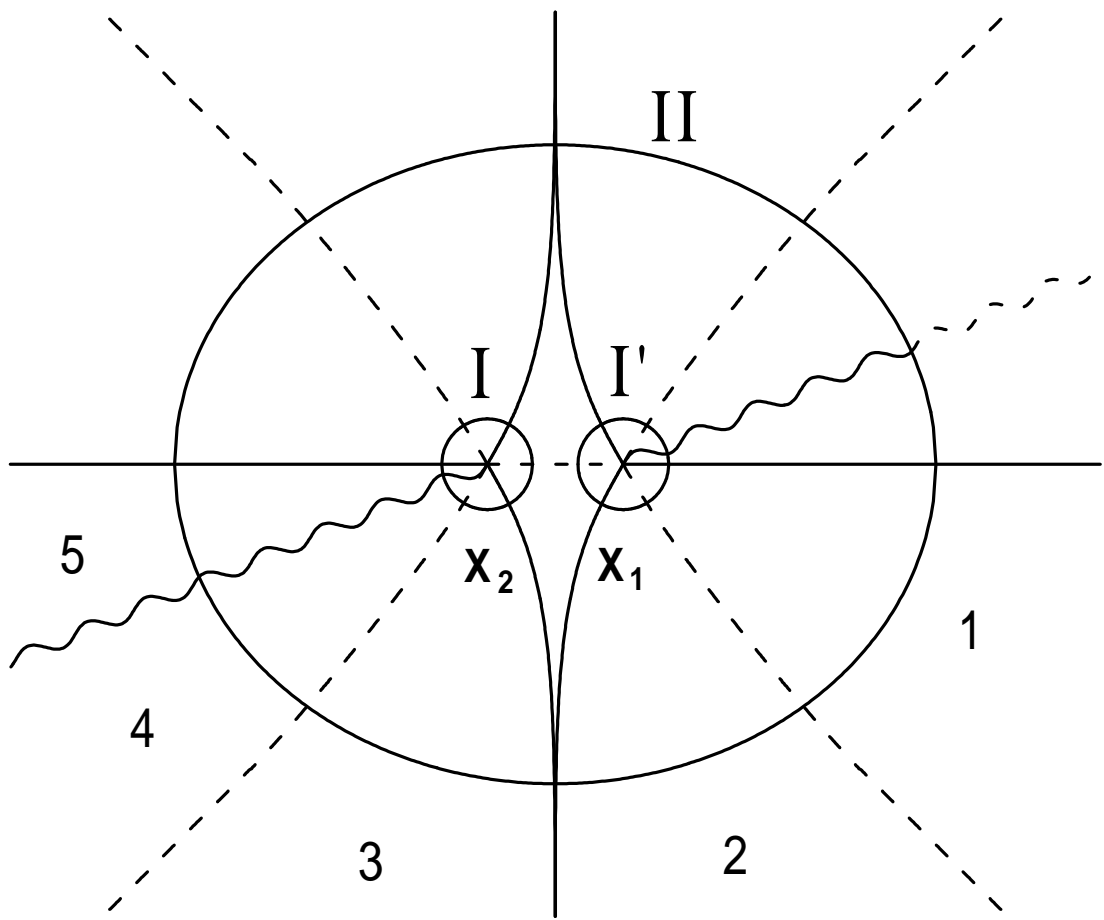
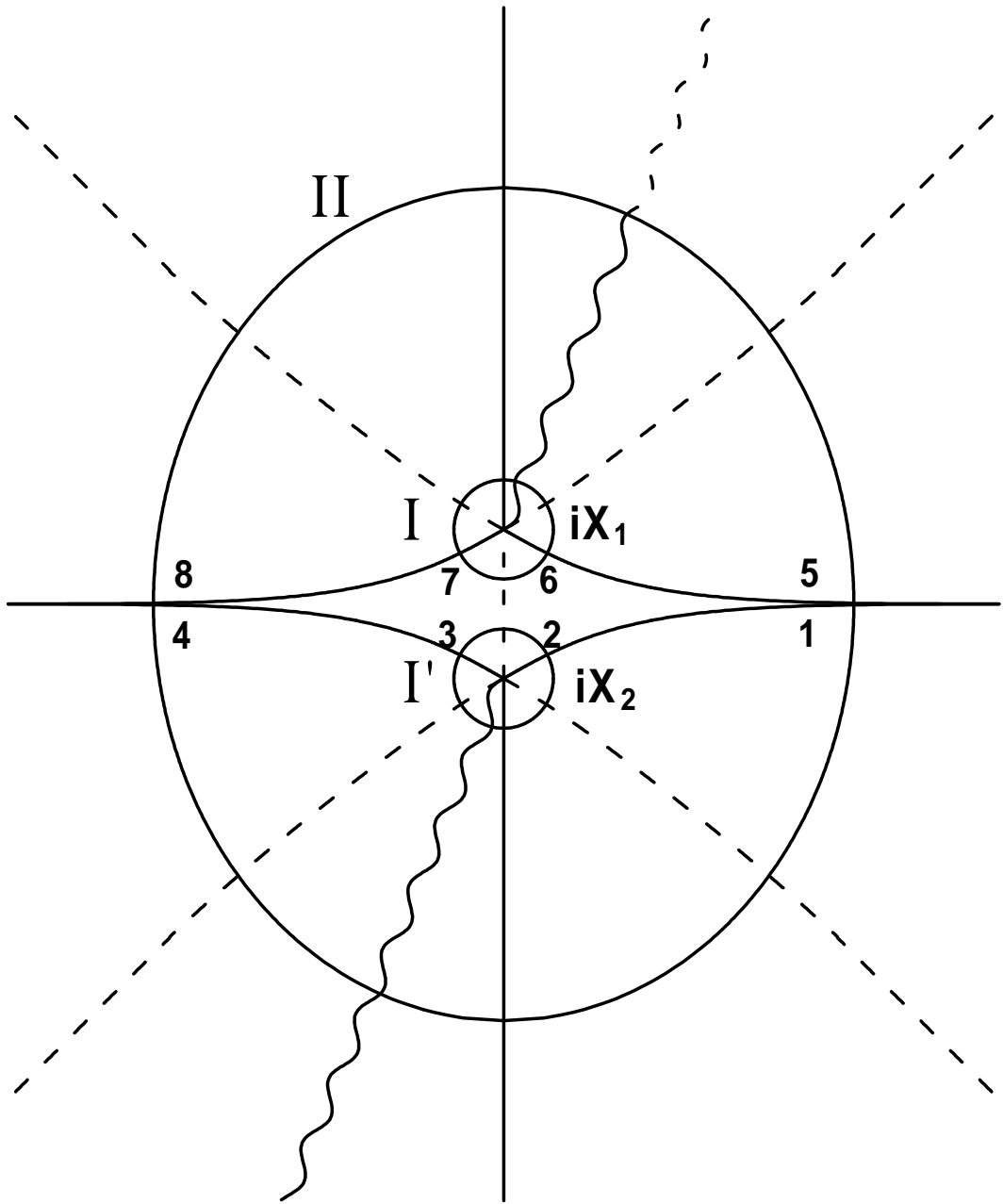
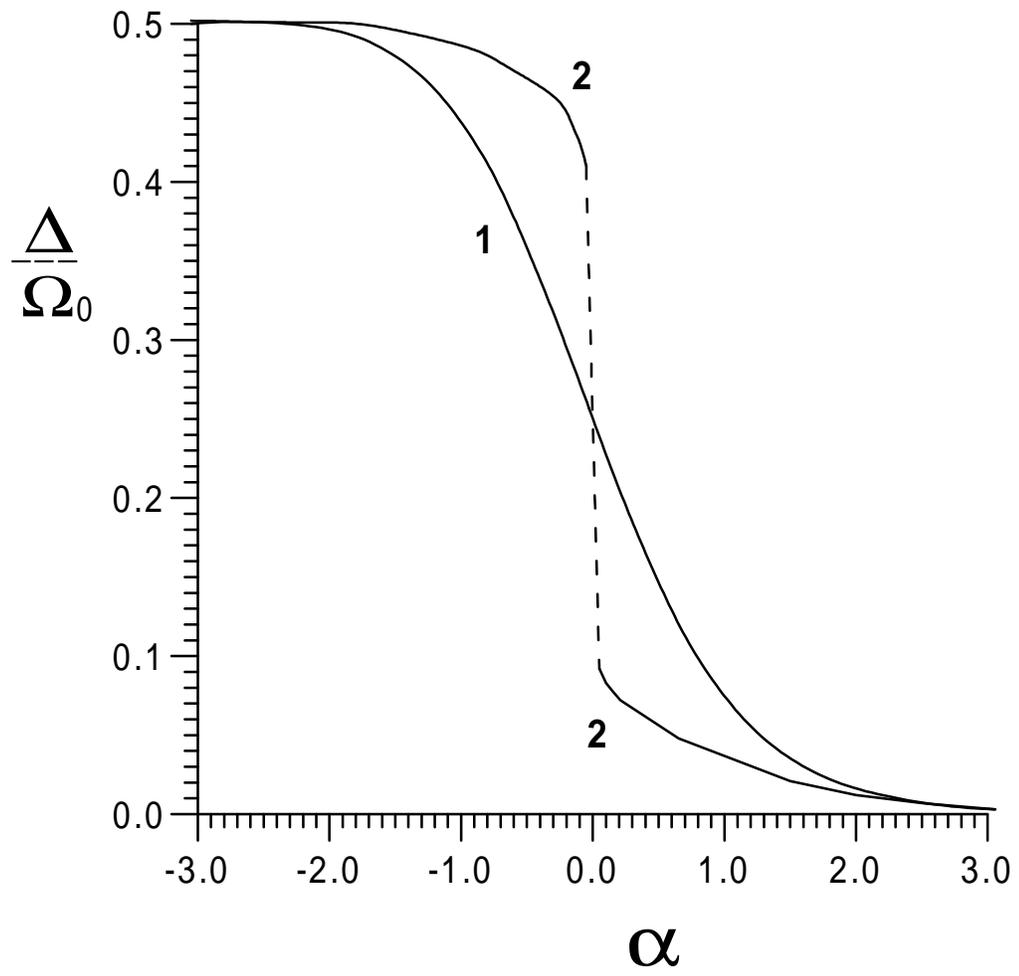
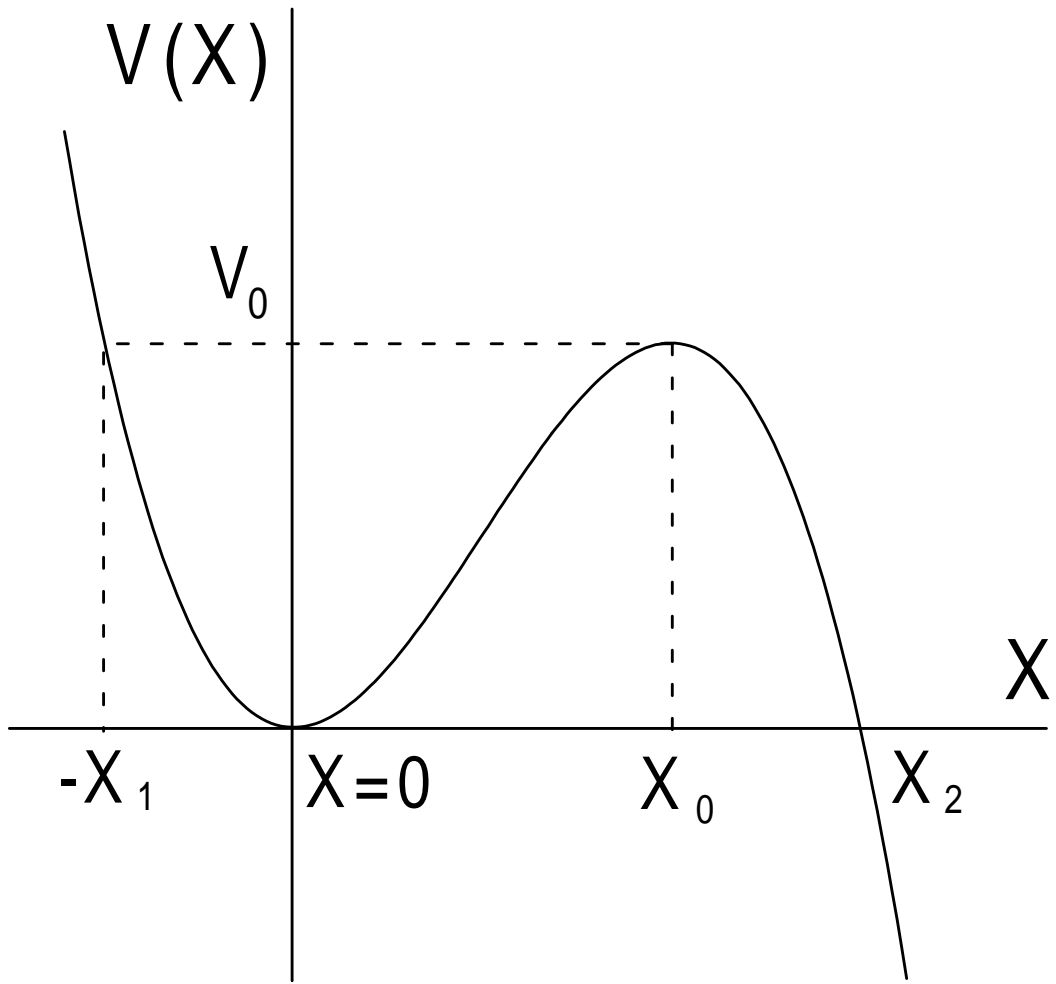
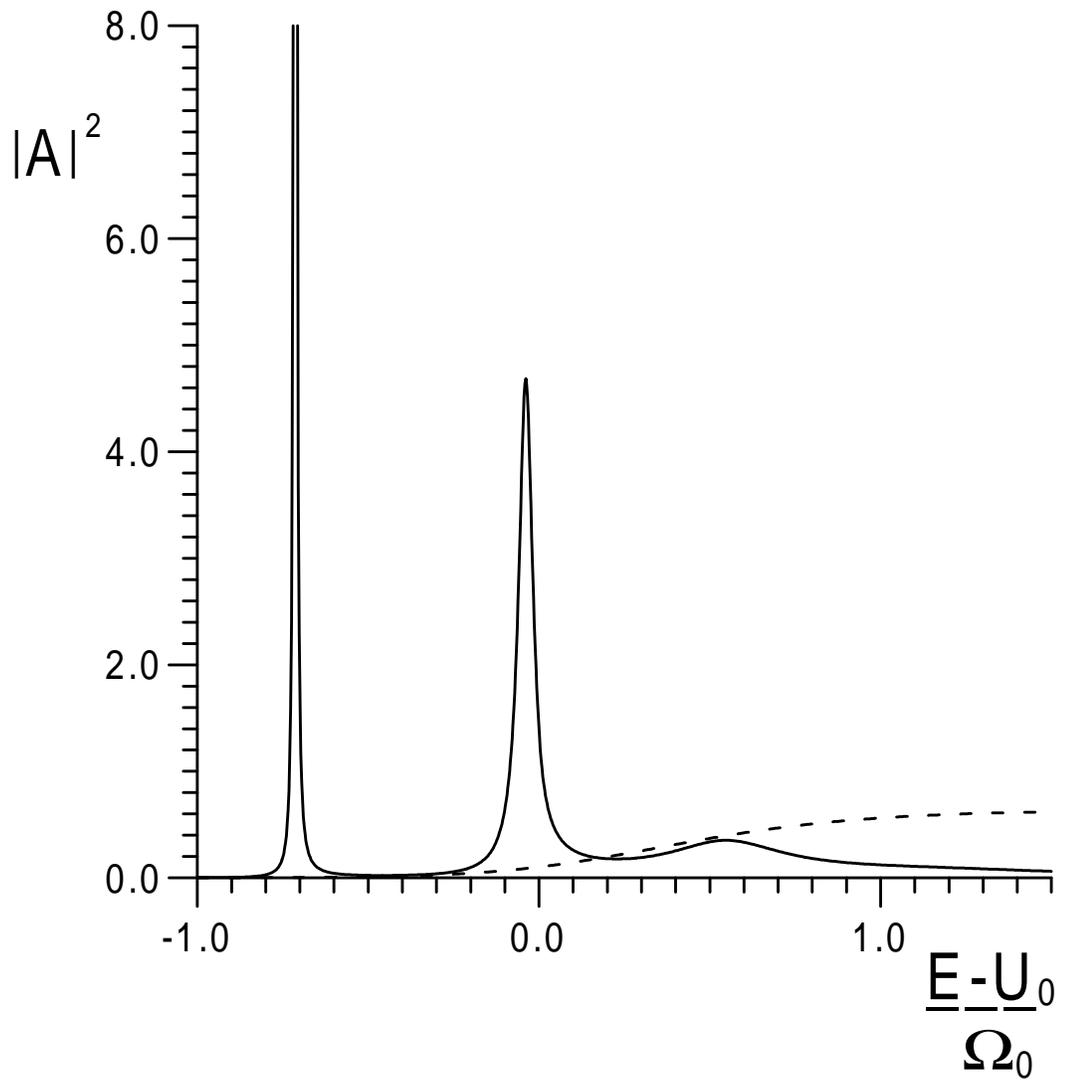
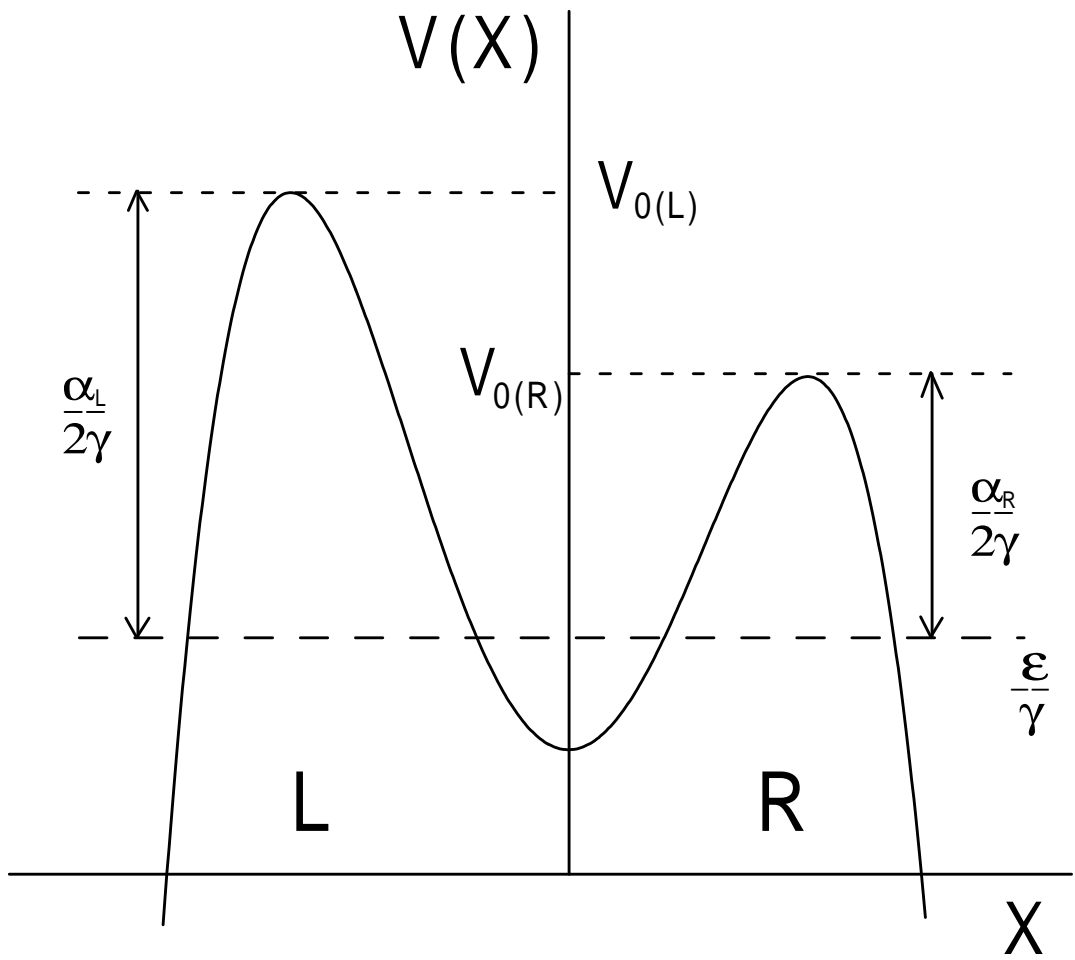
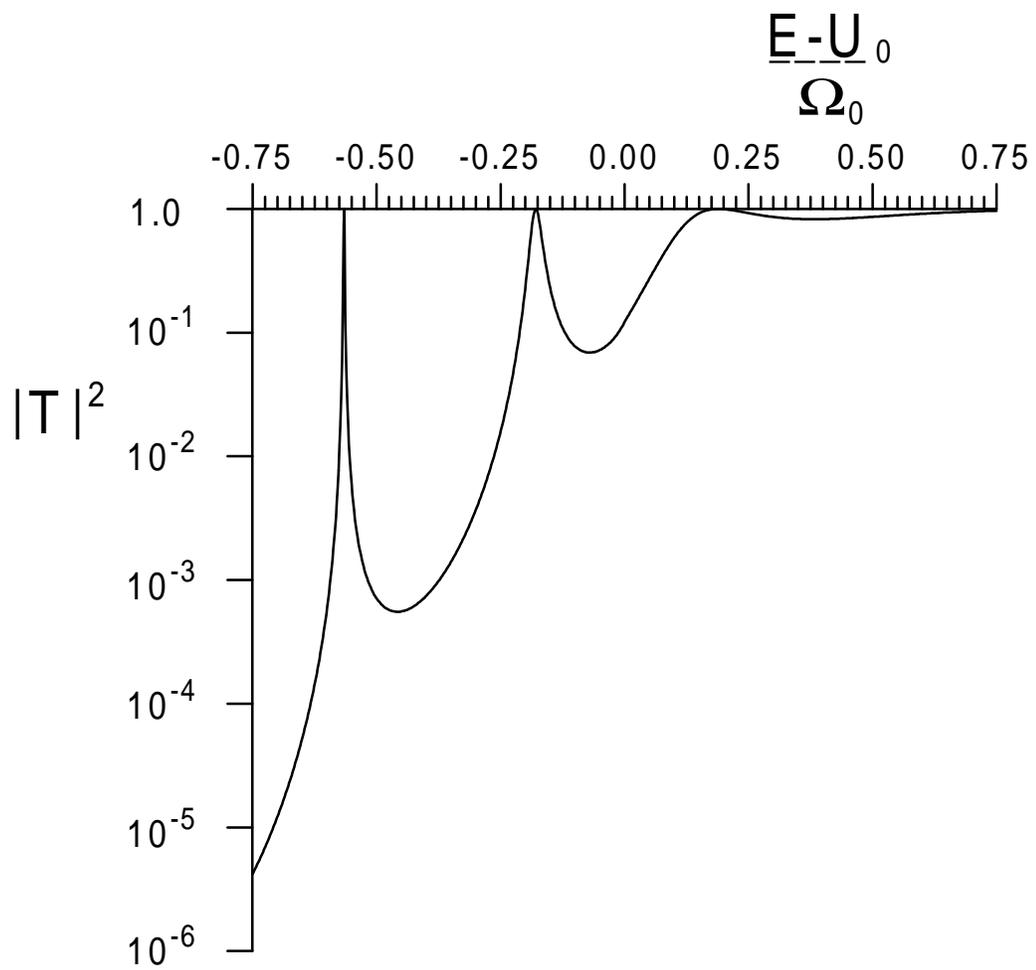
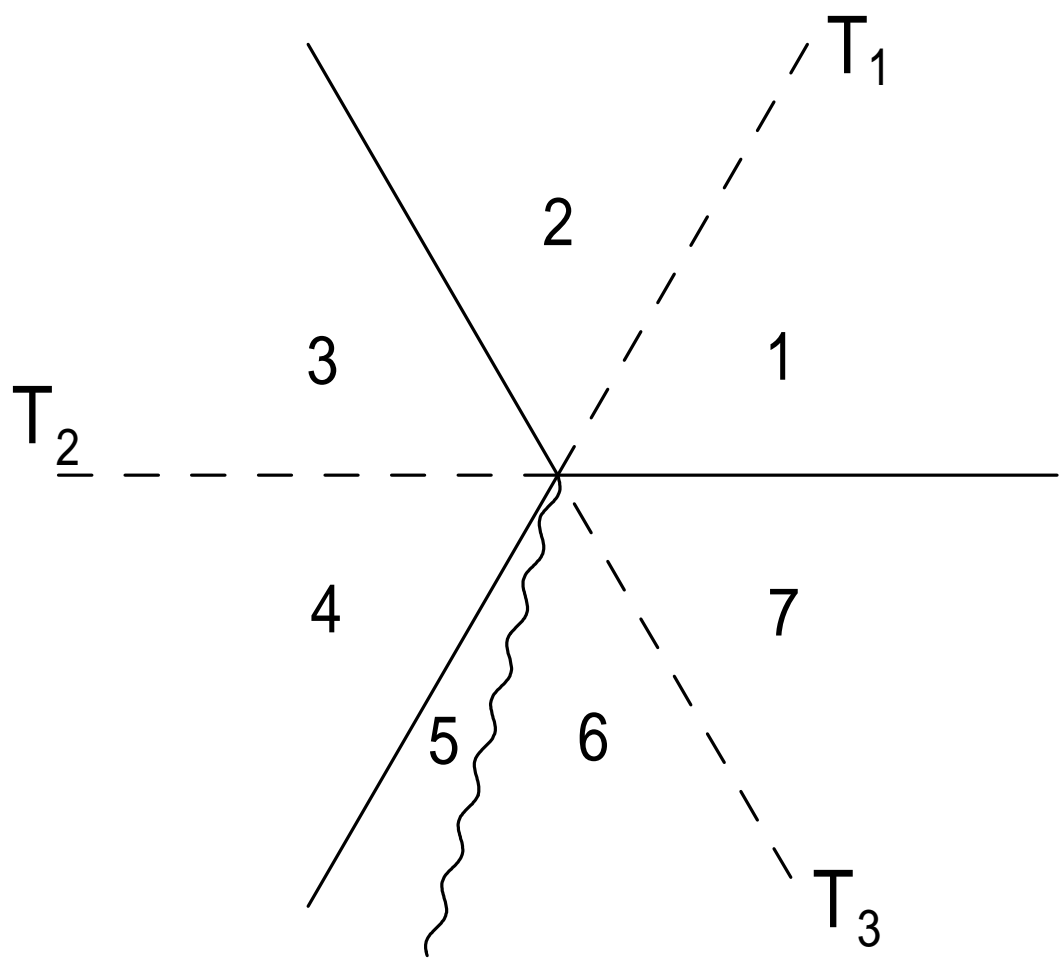
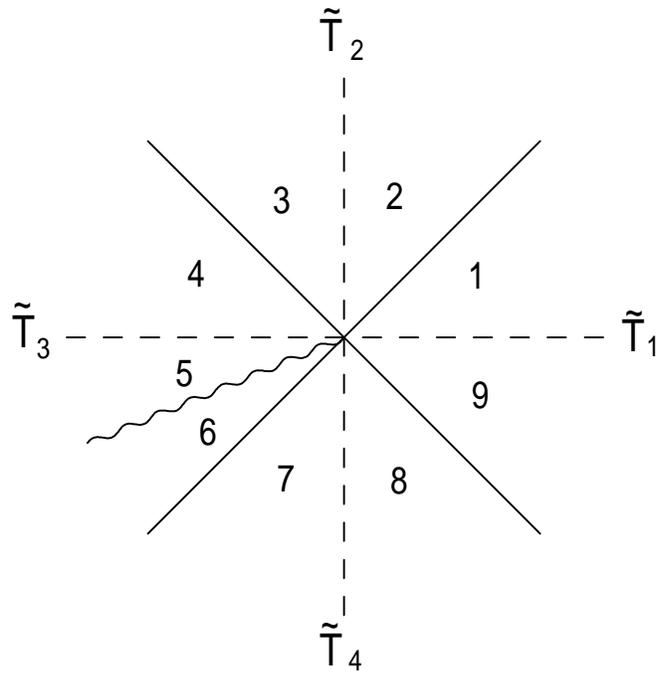
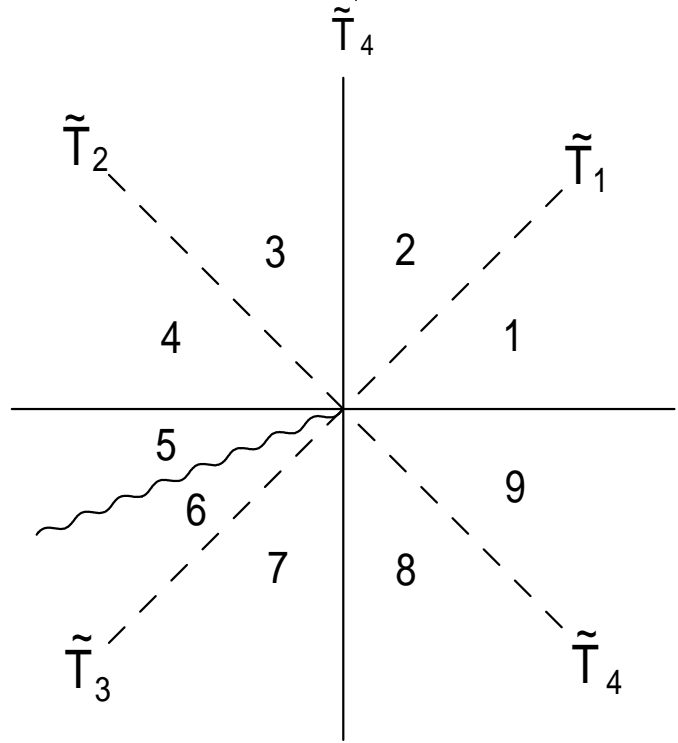
\end{document}